\documentclass[12pt]{article}
\usepackage{amsfonts}
\usepackage{amssymb}
\usepackage{amsbsy}
\usepackage{latexsym}
\evensidemargin=\oddsidemargin
\newcommand{\beq}{\begin{equation}}
\newcommand{\eeq}{\end{equation}}
\newcommand{\bea}{\begin{eqnarray}}
\newcommand{\eea}{\end{eqnarray}} 
\newcommand{\ba}{\begin{array}}
\newcommand{\ea}{\end{array}} 
\newcommand{\ds}{\displaystyle} 
\newcommand{\N}{{\cal N}} 
\newcommand{\NX}{{\cal N}^{(X)}} 
\newcommand{\Npsi}{{\cal N}^{(\psi)}} 
\newcommand{\D}{{\cal D}} 
\newcommand{\DX}{{\cal D}_{(X)}} 
\newcommand{\Dpsi}{{\cal D}_{(\psi)}} 
\newcommand{\bX}{{\mathbb X}}

\newcommand{\mA}{{\mathbb A}} 
\newcommand{\mF}{{\mathbb F}} 
\newcommand{\al}{\alpha}
\newcommand{\be}{\beta}
\newcommand{\la}{\lambda}
\newcommand{\e}{\epsilon}
\newcommand{\ha}{{\hat a}}
\newcommand{\hb}{{\hat b}}
\newcommand{\p}{\partial}
\newcommand{\Si}{\Sigma}
\newcommand{\dels}{\delta_{susy}}
\newcommand{\wt}{\widetilde}
\newcommand{\br}{\big\vert_{\p\Sigma}}
\newcommand{\rf}[1]{(\ref{#1})}
\begin{document}
\begin{flushright}
hep-th/0308201\\ 
USITP-03-07\\ 
\end{flushright}
\begin{center}
{\Large {\bf N=1 Worldsheet Boundary Couplings} 
\\[.2cm] 
{\bf and Covariance of non-Abelian Worldvolume Theory}\\} 
\vspace*{10mm} 
{\bf S. F. Hassan\footnote{\tt e-mail: fawad@physto.se}}\\
\vspace{2mm}
{\it Department of Physics, Stockholm University,\\[.1cm] 
AlbaNova University Centre, SE - 106 91 Stockholm, Sweden}
\vspace{.5cm}
\begin{abstract}
\noindent 
A systematic construction is given for N=1 open string boundary
coupling to Abelian and non-Abelian Dp-brane worldvolume fields, in
general curved backgrounds. The basic ingredient is a set of four
``boundary vectors'' that provide a unified description of boundary
conditions and boundary couplings. We then turn to the problem of
apparent inconsistency of non-Abelian worldvolume scalar couplings
(obtained by T-duality), with general covariance. It means that the
couplings cannot be obtained from a covariant action by gauge fixing
ordinary general coordinate transformations (GCT). It is shown that
the corresponding worldsheet theory has the same problem, but is also
invariant under certain matrix-valued coordinate transformations (MCT)
that can be used to restore its covariance. The same transformations
act on the worldvolume, leading to a covariant action. Then the
non-Abelian Dp-brane action obtained by T-duality corresponds to gauge
fixing the MCT and not GCT, hence the apparent incompatibility with
general covariance.
\end{abstract}
\end{center}
\setcounter{footnote}{0}
\baselineskip15pt
\section{Introduction and Summary}  \label{sInt}

The presence of non-Abelian scalar fields on the D-brane worldvolume
gives rise to new interactions investigated in
\cite{Taylor:1999gq,Taylor:1999pr,Myers:1999ps,Garousi:2000ea}, that
have widely been used since and lead to interesting phenomena. In
particular, in \cite{Myers:1999ps} the couplings of non-Abelian
scalars were obtained by T-dualizing D9-branes to Dp-branes. While
this is a consistent procedure, the resulting worldvolume action has a
puzzling feature, which is that it cannot be obtained from a general
covariant action on fixing a coordinate gauge, {\it e.g.}, the static
gauge. The reason is that some components of the non-Abelian scalars
which appear in a covariant description and are non-zero even in the
static gauge, do not show up in the action. Being matrices, they
cannot be gauged away by ordinary coordinate transformations. This
signals an apparent inconsistency of the theory with general
covariance which is not acceptable.

This apparent inconsistency of the non-Abelian worldvolume action with
general covariance cannot be attributed to a shortcoming in the
derivation. In fact, it arises within the regime of validity of the
procedure followed in \cite{Myers:1999ps}. Thus, while the missing
terms needed to restore compatibility with covariance are easy to
guess, we cannot simply introduce them into the action by hand as that
amounts to tampering with the outcome of a consistent calculation. The
resolution of the puzzle then requires understanding the origin of the
problem and finding a mechanism to restore the missing terms.

To achieve this, we study the string worldsheet boundary coupling to
worldvolume fields, which is the microscopic origin of the worldvolume
theory. We obtain the N=1 worldsheet boundary conditions and boundary
couplings to Abelian and non-Abelian worldvolume fields, and in general
backgrounds for general brane embeddings, which so far have not
been investigated satisfactorily. The problem is formulated in terms
of a set of four ``boundary vectors'', $\NX$, $\Npsi$, $\DX$ and
$\Dpsi$, that live on the worldsheet boundary and which have
components tangent and normal to the brane. The boundary conditions
follow from setting to zero appropriate projections of the boundary
vectors, while the surviving components describe the coupling of the
worldsheet boundary to worldvolume fields. This leads to a natural
description of worldsheet boundary couplings, consistent the general
covariance, supersymmetry and T-duality. It is shown that all
couplings to worldvolume scalars follow from shifts in the
coordinates, appropriately supersymmetrized and with the non-Abelian
features taken into account. The possibility of this interpretation in
the supersymmetric theory is intimately related to the correct
handling of the NS-NS 2-form gauge invariance on the boundary. Another
feature is that the gauge field is always shifted by a combination of
the scalars and the NS-NS 2-form field.

Having constructed the boundary action, we study how the boundary
coupling to non-Abelian Dp-brane follows from the D9-brane on
T-dualizing. The same procedure was applied in \cite{Myers:1999ps} to
the worldvolume theory. The scalar couplings in the worldsheet
boundary action obtained in this way share all features of the
corresponding couplings in the worldvolume theory, including the
apparent inconsistency with general covariance. However, the
worldsheet theory also develops a symmetry which resolves the problem:
Here covariance can be restored by the addition of terms that vanish
due to the boundary conditions, and hence do not change the content of
the theory. The vanishing terms are also generated by matrix-valued
coordinate transformations. It then follows that the worldvolume
theory should have a corresponding symmetry which involves shifting
ordinary coordinates by matrix valued functions. Correctly
implemented, this restores the covariance of the non-Abelian
worldvolume action.

The apparent inconsistency of the non-Abelian worldvolume theory
\cite{Myers:1999ps} with general covariance can now be understood
based on the picture that emerges from the worldsheet considerations:
$I$) Besides general coordinate transformations (GCT), the complete
worldvolume action is also invariant under a group of matrix-valued
coordinate transformations (MCT). Our worldsheet considerations probe
only a part of the MCT. Some aspects of such transformations have been
postulated and studied in \cite{DeBoer:2001uk,VanRaamsdonk:2003gj} in the
worldvolume theory. $II$) Now, the larger symmetry allows us to fix a
gauge using the MCT. This also fixes the GCT since they both act on
the same objects. $III$) The Dp-brane action that emerges from
T-duality is automatically in such a gauge which can thus be undone
only by an element of MCT and not GCT. Therefore, if the existence of
the MCT is not taken into account, one cannot relate this action to a
covariant one through GCT alone, hence the apparent inconsistency
between the two. $IV$) The worldsheet analysis provided us with the
right element of MCT to undo the gauge fixing. The admissibility of
this matrix-valued transformations follows from the worldsheet theory
and is not obvious from the known structure of the worldvolume action
in \cite{Myers:1999ps}. This summarizes the solution to the problem of
covariance of the non-Abelian worldvolume theory.

The paper is organized as follows: In section 2 we introduce a set of
four ``boundary vectors'' and express the boundary term in the
variation of the N=1 supersymmetric worldsheet action in terms of
them. In section 3, we briefly review the basics of the covariant
description of brane geometry, with emphasis on the worldvolume
scalars and their structure in the static gauge. We then obtain the
covariant form of Dp-brane boundary conditions in terms of the
boundary vectors and examine their consistency with T-duality. In
section 4, we construct worldsheet boundary couplings in terms of the
boundary vectors and examine their behaviour under T-duality. The idea
is first implemented in the bosonic theory and then generalized to N=1
Abelian and non-Abelian cases. The boundary couplings to scalars are
interpreted as appropriate coordinate shifts in all the three cases.
This crucially depends on the manifest NS 2-form gauge invariance of
the couplings. The behaviour of the boundary action under T-duality
suggests the possibility of coordinate shifts by non-Abelian matrices
to restore compatibility with general covariance. In section 5, we
turn to the issue of the compatibility of non-Abelian scalar couplings
with general covariance in the worldvolume theory and its resolution
based on our worldsheet considerations. As a test, the idea is first
applied to the Abelian worldvolume theory and shown to work. We then
consider the non-Abelian case and obtain the general covariant form of
the scalar couplings. We also outline the general picture that emerges
from the analysis and which puts our results in perspective. The
implication for the formulation of D-brane interactions in terms of
Clifford multiplication is also discussed. The conclusions are
summarized in section 6. Appendix A explains our supersymmetry
conventions and Appendix B contains the index conventions.

\section{Boundary Variation and Boundary Vectors} \label{sBVBV}

In this section we consider NSR open strings with N=1 worldsheet
supersymmetry in general metric $G_{MN}(X)$ and antisymmetric tensor
field $B_{MN}(X)$ backgrounds. The boundary terms that arise on
varying the action are written in terms of a set of ``boundary
vectors'' which provide a natural realization of supersymmetry. They
will play an important role in later sections. 

\subsection{Boundary variations} \label{ssBVar}

In the absence of boundary interactions, the open string worldsheet
action with N=1 supersymmetry is (see the appendix for conventions),  
\beq 
S=\frac{1}{2}\int_\Sigma d^2\sigma\int d\theta^+\int d\theta^- 
E_{MN}(\bX)D_+\bX^M D_-\bX^N + \frac{i}{4}\int_{\p\Sigma}d\tau 
B_{MN}(\psi^M_+\psi^N_+ + \psi^M_-\psi^N_-) \,.
\label{actionSF}
\eeq
Here, $\bX$ is the N=1 superfield and $E=G+B$. This form of the action
is known to be correct for constant backgrounds and no further
modifications arise in space-time dependent backgrounds. The added
boundary term in (\ref{actionSF}) has been noted in \cite{
Borlaf:1996na,Govindarajan:2000ef,Bachas:2001vj,Albertsson:2002qc}
but its presence can be argued for on very general grounds: It is
needed to cancel a similar term that is hidden in the superspace part
so that, in terms of the component fields, the action $S$ does not
contain boundary terms and has the same form as the closed string
action. Such a boundary term, if present, would have meant that the
bulk field $B_{MN}$ couples differently to open and closed strings,
which should not be the case. Then, in terms of component fields and
after integrating out the auxiliary field, one obtains,
\bea 
S &=&\frac{1}{2} \int_\Si d^2\sigma
\Big[ \p_+X^M E_{MN} \p_-X^N +i\psi^M_+ G_{MN}\nabla_-\psi^N_+
+i\psi^M_-G_{MN}\nabla_+\psi^N_- \nonumber\\ 
&&\qquad \qquad \qquad \qquad\qquad \qquad\qquad \qquad
+\,\frac{1}{2}\,\psi_+^M\psi_+^N\psi_-^K\psi_-^L R_{MNKL}(E) \Big]\,.
\label{actionCom} 
\eea  
The covariant derivatives are given by $\nabla_{\pm}\psi^N_\mp = 
\p_\pm\psi^N_\mp\,+\,(\Gamma^N_{LK}\,\mp\,\frac{1}{2} H^N_{\,\,\,LK}) \p_\pm 
X^L\psi^K_\mp$, and $ R_{MNKL}(E)$ is the curvature tensor associated
with $\nabla_{+}$. 

Boundary terms arise in the variation $\delta S$ of the action when
deriving the equations of motion, as well as in its supersymmetry
variation $\delta_{susy} S$. Both these should vanish by the same 
set of boundary conditions. The bulk term in $\delta S$ yields the
equations of motion, hence we retain only the boundary terms, 
\bea
\delta S &=& \frac{1}{2}\int_{\p\Sigma} d\tau \Big[\delta X^L 
\Big( E_{LM}\p_-X^M -\p_+X^M E_{ML}+i\psi_-^M G_{MK}\Omega^K_{LN}(E) 
\psi^N_- \nonumber\\
&&\qquad\qquad\qquad - i \psi_+^M G_{MK}\Omega^K_{LN}(E^T)\psi^N_+ \Big)
+ i \psi_-^M G_{ML}\delta \psi^L_-
- i \psi_+^M G_{ML}\delta \psi^L_+ \Big] \,.
\label{dSraw}
\eea
Naively one may regard $\delta X^L$ and $\delta\psi^L_\pm$ as
independent variations and set the associated boundary terms to 
zero separately. For constant backgrounds this will lead to the
correct boundary conditions. For $X$-dependent backgrounds, this turns
out to be inconsistent with both worldsheet supersymmetry and with
T-duality. To get the correct boundary conditions, we reorganize the
terms in $\delta S$ writing them as a sum of bosonic and fermionic
boundary terms, $(BBT)$ and $(FBT)$,
\beq
\delta S =\frac{1}{2}\int_{\p\Sigma}d\tau \Big[(BBT)+i(FBT)\Big]\,,
\label{action}
\eeq 
where, 
\bea
(BBT)&=&\delta X^L\Big(E_{LM}\p_-X^M - E^T_{LM} \p_+ X^M \nonumber\\[.1cm]
&& -i\psi_-^M \p_M E_{LN}\psi^N_- +i\psi_+^M\p_M E^T_{LN}\psi^N_+ 
-i\eta \psi_+^M\p_L E_{MN}\psi_-^N \Big) \,\br \,,
\label{BBT}\\[.1cm]
(FBT)&=& \psi^M_- G_{MN}\delta\psi^N_- -\psi^M_+ G_{MN}\delta\psi^N_+
\nonumber\\[.1cm]
&& + \delta X^L \Big(\frac{1}{2}\left[\psi_-^M \p_L B_{MN}\psi^N_- 
+\psi_+^M\p_L B_{MN}\psi^N_+ \right] 
+\eta \psi_+^M\p_L E_{MN}\psi_-^N \Big) \,\br \,.
\label{FBT}
\eea
This decomposition may seem arbitrary but in fact it is fixed by
worldsheet supersymmetry as well as by consistency with the constant
background case. The parameter $\eta$, which is so far arbitrary (as
it cancels between \rf{BBT} and \rf{FBT}), will be required to take
values $\pm 1$. These can be assigned independently at both ends of
the open string, $\sigma=0,\pi$. From the example of constant
backgrounds we know that if $\eta$ is assigned the same value at both
ends, we are in the Ramond (R) sector while opposite $\eta$
assignments at $\sigma=0$ and $\sigma=\pi$ lead to the Neveu-Schwarz
(NS) sector.

In general under a supersymmetry transformation the action changes by
a boundary term $\dels S$. For boundary conditions to be consistent
with N=1 worldsheet supersymmetry, they should also imply the vanishing
of $\dels S$. After some manipulations, one finds that the variation of
\rf{actionSF} under a supersymmetry transformation is 
\beq
\delta_{susy} S =\frac{1}{2}\int_{\p\Sigma}d\tau \Big[ (BBT) + 
(i+2) (FBT) \Big]\,.
\label{susydS}
\eeq 
Here $(BBT)$ and $(FBT)$ are still given by \rf{BBT} and \rf{FBT} with 
$\delta X$ and $\delta\psi_\pm$ replaced by the corresponding
supersymmetry variations $\delta_{susy} X$ and $\delta_{susy}\psi_\pm$
given in the appendix. We have also assumed that the left and the
right supersymmetry transformation parameters are related by
$\e^-=\eta\e^+$ as follows from restricting to constant backgrounds. 

Comparing this with $\delta S$ in \rf{action} it is now clear that
boundary conditions should set $(BBT)$ and $(FBT)$ to zero
independently. This is how worldsheet supersymmetry justifies the
split in \rf{action}.

\subsection{Boundary ``vectors''} \label{ssBVec}

To develop a systematic description of the boundary conditions and
boundary couplings, we now introduce two bosonic quantities $\NX$,
$\DX$ and two fermionic ones $\Npsi$, $\Dpsi$ that live on the
worldsheet boundary, 
\bea
\NX_L &=& E_{LM}\p_-X^M - E^T_{LM} \p_+ X^M \nonumber\\[.1cm]
&&-i\psi_-^M \p_M E_{LN}\psi^N_- +i\psi_+^M\p_M E^T_{LN}\psi^N_+ 
-i\eta \psi_+^M\p_L E_{MN}\psi_-^N \,\br \,,
\label{B} \\[.1cm]
\Npsi_L &=& E_{LN}\psi_-^N - \eta E^T_{LN}\psi_+^N \,,\br\,,
\label{F} \\[.1cm]
\DX^L &=& \p_\tau X^L \,\br \,, 
\label{DX}\\[.1cm]
\Dpsi^L &=& \psi_-^L + \eta\psi_+^L \, \br \,. 
\label{Dpsi}
\eea 
We will refer to these as ``boundary vectors'' (even though $\NX$ is
not really a vector under general coordinate transformations). As the
nomenclature (and the form) suggests, they will be associated with
Neumann and Dirichlet boundary conditions.

To convince ourselves that these are the natural objects to work with,
we look at the behaviour of boundary vectors under supersymmetry.
Using the supersymmetry transformations of $X^M$ and $\psi^M_\pm$ 
and taking the left and right supersymmetry transformation parameters
to be related by $\e^-=\eta\e^+$, one can check that the boundary
vectors have very simple transformation properties under the N=1
worldsheet supersymmetry,  
\beq
\begin{array}{ccc}
\dels\Npsi_M=-i\e^-\NX_M\,,&\qquad&\dels\NX_M=-2\e^-\p_\tau\Npsi_M\,,
\\[.3cm]
\dels\Dpsi^M=-2i\e^-\DX^M\,,&\qquad&\dels\DX^M=-\e^-\p_\tau\Dpsi^M\,.
\end{array}
\label{BVSUSY}
\eeq

We can now express the boundary term in the variation of the action in
terms of the boundary vectors. Note that under variations $\delta X^L$
and $\delta\psi^L_\pm$, 
\bea
\delta \Npsi_L &=& E_{LN}\delta\psi_-^N - \eta E^T_{LN}\delta\psi_+^N +
\delta X^K\left(\p_K E_{LN}\psi_-^N-\eta\p_K
E^T_{LN}\psi_+^N\right)\,\br\,,\label{deltaF} \\[.1cm]
\delta \Dpsi^L  &=& \delta\psi_-^L + \eta\delta \psi_+^L \, \br \,. 
\label{deltaDpsi}
\eea
Then, in terms of the boundary vectors, the expressions $(BBT)$
\rf{BBT} and $(FBT)$ \rf{FBT} take the simple forms,  
\bea
(BBT)&=&\delta X^L \NX_L \,,
\label{BBTB}\\[.1cm]
(FBT)&=& \frac{1}{2}\,\Dpsi^L\delta\Npsi_L
-\frac{1}{2}\,\delta\Dpsi^L \Npsi_L \,,
\label{FBTF}
\eea
and the boundary terms in the variation of the action \rf{action}
become 
\beq
\delta S =\frac{1}{2}\int_{\p\Sigma}d\tau \Big[ \delta X^L \NX_L + 
\frac{i}{2}\,\Dpsi^L\delta\Npsi_L -\frac{i}{2}\,\delta\Dpsi^L \Npsi_L 
\Big]\,.
\label{actionBV}
\eeq
The boundary conditions now follow in a straightforward way from the 
requirement that $(BBT)$ and $(FBT)$ vanish independently.

As an example, let us consider the simplest case of D9-branes. For
constant backgrounds, the boundary conditions have been known for a
long time \cite{Callan:1988wz}. In space-time dependent backgrounds,
the problem was considered in \cite{Borlaf:1996na}. It was later
revisited in \cite{Albertsson:2002qc,Albertsson:2001dv} where a
general parameterization of the boundary conditions was investigated
by studying the classical N=1 superconformal algebra. In our approach,
for open strings on a D9-brane, the Neumann boundary conditions follow
directly from the the vanishing of the boundary variation
\rf{actionBV} as $\NX_L = 0$ and $\Npsi_L = 0$, in agreement with 
\cite{Albertsson:2002qc}. We write these explicitly for later
reference,   
\bea
&E_{LM}\p_-X^M\!\!-\!\!E^T_{LM} \p_+ X^M\!\! 
-\!i\psi_-^M \p_M E_{LN}\psi^N_-\!+\!i\psi_+^M\p_M E^T_{LN}\psi^N_+ 
\!-\!i\eta \psi_+^M\p_L E_{MN}\psi_-^N \br = 0\,,&
\label{D9BN}\\[.2cm]
&E_{LN}\psi_-^N - \eta E^T_{LN}\psi_+^N \br =0\,.&
\label{D9FN}
\eea
As usual, the variations $\delta\psi_\pm$, $\delta X$ are restricted
to the class of functions that satisfy the boundary conditions so that
$\Npsi_L = 0$ also implies $\delta\Npsi_L=0$, leading to $\delta S=0$. 

\section{Dp-brane Boundary Conditions} \label{sBC}

We start this section with a brief review of the covariant description
of Dp-branes as embedded submanifolds in space-time. These geometrical
notions are used in the rest of the paper. We then obtain the Dp-brane
boundary conditions which take a particularly simple form in terms of
the boundary vectors. The boundary conditions are then shown to be
consistent with T-duality.

\subsection{Covariant description of Dp-branes} \label{ssCov}

Let coordinates $\xi^\al$ ($\al=0,1,\cdots,p$) parameterize the
Dp-brane worldvolume. The embedding of the worldvolume as a
hypersurface in space-time is then described by the functions
$X^M(\xi)$. The metric induced on the worldvolume is
$g_{\al\be}=G_{MN}\p_\al X^M\p_\be X^N$. The tangent and normal
bundles to the brane are spanned by basis vectors $\p_\al X^M\p_M$ and
$a^M_\ha\p_M$, respectively. $\ha=p+1,\cdots,9$ is a flat normal
bundle index raised and lowered by the flat metric $\delta_{\ha\hb}$
and $a_M^\ha a^M_\hb=\delta^\ha_\hb$. Worldvolume and space-time
indices are raised and lowered by the corresponding metrics
$g_{\al\be}$ and $G_{MN}$, respectively. The orthogonality of tangent
and normal vectors implies
\beq
a^M_\ha G_{MN}\p_\al X^N = a_{\ha N}\p_\al X^N = 0 \,.
\label{adX} 
\eeq 
The three metrics $g_{\al\be}$, $G_{MN}$ and $\delta_{\ha\hb}$ are
related via 
\beq
G^{MN}=a^M_\ha\delta^{\ha\hb}a^N_\hb + \p_\al X^M g^{\al\be}\p_\be X^N\,.
\label{metrics}
\eeq
Space-time vectors $V^M$ and $V_M$ have projections along the tangent
and normal directions given by, 
\beq
V_\al = V_M\p_\al X^M\,,\quad V_\ha = V_M  a^M_\ha\,;\qquad 
V^\al = V^M G_{MN}\p_\be X^N g^{\be\al}\,,\quad V^\ha = V^M  a_M^\ha\,. 
\label{V}
\eeq
Using these one can verify that all indices can be consistently raised
or lowered by the corresponding metrics $G_{MN}$, $g_{\al\be}$ and
$\delta_{\ha\hb}$. Conversely, a vector $V^M$ on the hypersurface can
be reconstructed in terms of its tangential and normal components,
\beq
V^M = a^M_\ha V^\ha + \p_\al X^M V^\al \,,
\label{Vtn}
\eeq
similarly, for $V_M =G_{MN}V^N$. 

D-branes contain gauge fields and transverse scalars fields living on
their worldvolumes. Since these are vector fields intrinsically
tangent (for gauge fields) and normal (for scalars fields) to the
brane, they naturally have components $\mA_\al$ and $\Phi^\ha$,
respectively. However, one can also work in terms of the corresponding
space-time components,
\beq
A_M = G_{MN}\p_\al X^N g^{\al\be} \mA_\be\,,\qquad
\phi^M = a^M_\ha\Phi^\ha\,.
\label{wv-st}
\eeq
Equation \rf{adX} insures that $A_M$ is tangential and $\phi^M$ is
normal to the D-brane.

When performing T-dualities or computing scattering amplitudes, the
manipulations involve components of $A$ and $\phi$ and often one does
not directly deal with $\mA$ and $\Phi$. For this reason we will use
different notations for these intrinsically worldvolume objects and
their space-time projections. Throughout the paper, the indices on
these fields and other quantities are raised and lowered by the
corresponding metrics, $G_{MN}$, $g_{\al\be}$ and $\delta_{\ha\hb}$.
 
D-brane worldvolume actions are often computed in the static gauge to
which the above description can be specialized. In static gauge, the
embedding functions $X^M(\xi^\al)$ are given by,  
\beq 
X^\mu=\xi^\mu\,\, (\mu=0,1,\cdots,p)\,,\qquad X^i=const \,\,
(i=p+1,\cdots, 9).  
\label{sg}
\eeq 
One can verify from \rf{adX} that in this gauge, $a^\ha_\mu=0$ and the
normal frame is fully spanned by $a^\ha_i$. However, For a general
metric $G_{MN}$, generically all components of $a^M_\ha=G^{Mi}
a^\ha_i$ are non-zero. Clearly the choice of static gauge affects only
$\phi^M$ but not $\Phi^\ha$ which contains the actual degrees of
freedom. In particular \rf{wv-st} implies that in the static gauge the
only non-zero components of $\phi_M$ are the $\phi_i$, while
generically, all components of $\phi^M$ are non-zero. This simple fact
is important in understanding the covariance of scalar field couplings
in D-brane worldvolume actions and hence is emphasized here. To
summarize, in the the static gauge,
\beq
\left. \ba{c}
a_\ha^M=\{a_\ha^\mu\,,a_\ha^i\}  
\\[.3cm]
a^\ha_M=\{a^\ha_\mu=0\,,a^\ha_i\}
\ea \right.
\;\;  \Rightarrow \;\;
\ba{c}
\phi^M=\{\phi^\mu\,,\phi^i\}
\\[.3cm]
\phi_M=\{\phi_\mu=0\,,\phi_i\} 
\ea
\label{aphisg}
\eeq
All this remains unchanged if \rf{sg} is slightly generalized to 
\beq 
X^\mu=X^\mu(\xi^\al)\,\, (\mu, \al =0,1,\cdots,p)\,,\qquad X^i=const\,\,
(i=p+1,\cdots, 9).  
\label{gsg}
\eeq 
For us the difference between the two will be immaterial and both will
be called the {\it static gauge}. 

\subsection{Dp-brane boundary conditions} \label{ssDpBC}

We now turn to Dp-brane boundary conditions implied by the vanishing
of $(BBT)$ and $(FBT)$ in \rf{BBTB} and \rf{FBTF}, first discussing
the general covariant case and then going to the static gauge.  

Consider the brane embedding $X^L(\xi)$. By definition, the boundary
of the worldsheet parameterized by $\tau$ is confined to the brane,
$X^L\br = X^L(\xi(\tau))$. As a result, the variations $\delta X^L\br$
are entirely tangent to the brane and have no components in directions
normal to it. This is the Dirichlet boundary condition on the
bosonic field $X^L$ that can be expressed in various equivalent ways,
\beq
{\rm Dirichlet}:\quad \delta X^L\br = \p_\al X^L \delta \xi^\al\,,
\quad {\rm or}\quad  a^\ha_L \delta X^L\br =0\,,
\quad {\rm or}\quad a^\ha_L \p_\tau X^L\br =0\,.
\label{DpBD}
\eeq
Here, the first equation along with \rf{adX} implies the second
equation and the third follows from the fact that the worldsheet
boundary confined to the brane is parameterized by $\tau$, so that
$\p_\tau X^L\br=\p_\al X^L\p_\tau \xi^\al$. Since the $\delta\xi^\al$
are arbitrary, the vanishing of the bosonic boundary term $(BBT)$
in \rf{BBT} or \rf{BBTB} amounts to $\p_\al X^L \NX_L\br=0$, which is 
the bosonic Neumann boundary conditions for Dp-branes,     
\bea
{\rm Neumann}: && \p_\al X^L \Big( E_{LM}\p_-X^M - E^T_{LM} \p_+ X^M 
\nonumber\\[.1cm] &&\qquad
-i\psi_-^M \p_M E_{LN}\psi^N_- +i\psi_+^M\p_M E^T_{LN}\psi^N_+ 
-i\eta \psi_+^M\p_L E_{MN}\psi_-^N \Big)\br = 0 \,.
\label{DpBN}
\eea 
To obtain the corresponding fermionic boundary conditions note that
the N=1 supersymmetry variation of $X^L$ is $\delta_{susy} X^L =
-(\e^- \psi_-^L + \e^+ \psi_+^L)$. On the boundary, $a^\ha_
L\delta_{susy}X^L\br=0$ by \rf{DpBD} and  $\e^+=\eta\e^-$, leading to,  
\beq
{\rm Dirichlet}:\qquad a^\ha_L\, (\psi_-^L + \eta\psi_+^L)\br
=  a^\ha_L \,\Dpsi^L =0\,. 
\label{DpFD}
\eeq
Since, by \rf{Vtn}, one can alway write $\Dpsi^L=\Dpsi^\ha a^L_\ha +
\Dpsi^\al \p_\al X^L$, then the vanishing of the fermionic boundary
term $(FBT)$ in \rf{FBTF} only requires the tangential component of
$\Npsi_L$ to vanish, $\p_\al X^L \Npsi_L\br=0$.  This is the fermionic
Neumann boundary condition for Dp-branes, 
\beq
{\rm Neumann}: \qquad \p_\al X^L \Big(E_{LN}\psi_-^N - 
\eta E^T_{LN}\psi_+^N \Big)\br = 0 \,.
\label{DpFN}
\eeq   
Equations \rf{DpBD}-\rf{DpFN} complete the Dp-brane boundary
conditions with N=1 worldsheet supersymmetry in general backgrounds
and for general embedding. 

We summarize the Dp-brane Neumann and Dirichlet boundary conditions in
terms of the boundary vectors \rf{B}-\rf{Dpsi}, 
\beq
\begin{array}{rcr}
\NX_\al\equiv\p_\al X^M\NX_M=0\,,&\qquad &\Npsi_\al\equiv\p_\al X^M
\Npsi_M=0 \,,   \\[.2cm]
\DX^\ha\equiv a^\ha_M\DX^M=0  \,,& \qquad & \Dpsi^\ha\equiv 
a^\ha_M\Dpsi^M=0 \,.
\end{array}
\label{BVBC}
\eeq
While the boundary vectors have enabled us to write elegant
expressions for the boundary conditions, their usefulness extends
beyond this. We will see in the next sections that the couplings of
open strings to the worldvolume fields are naturally given in terms of
the boundary vector components that are not set to zero by the
boundary conditions.

Before proceeding further, we make some observations that contrast
and compare boundary conditions in the N=1 theory to the bosonic
theory, or the case with constant backgrounds: 
\begin{enumerate}
\item Under open-closed string duality ($\sigma\leftrightarrow\tau$
accompanied by $E_{MN}\rightarrow -E_{MN}$ and $\psi^M_-\rightarrow
i\psi^M_-$, to keep the action invariant), $\NX_\al$ does not go over
to the closed string canonical momentum $P_\al$. The two differ by
terms that are bilinear in worldsheet fermions and are proportional to
$\p_L E_{MN}$.
\item The ``boundary vector'' $\NX_L$ does not transform covariantly
under space-time general coordinate transformations due to the
non-covariance of its fermionic content (so it is not really a
vector). This is because the structure of $\NX_L$ is constrained by
worldsheet supersymmetry which combines vectors $\psi^M_\pm$ with the
coordinate $X^M$ in the same superfield. But the final expressions
are always covariant. 
\item Contrary to some statements in the literature, mere consistency
of Dp-brane boundary conditions with N=1 worldsheet supersymmetry does
not constrain the background fields in any way. Such constrains have
to come from separate stability considerations and space-time
supersymmetry.
\end{enumerate}

The boundary conditions can now be specialized to the often used
case of static gauge. The resulting expressions will be needed in the
remaining sections to understand the relation between T-duality and
covariance of the worldvolume actions. Break up the space-time
coordinates $X^M$ into two sets $X^\mu$ (for $\mu=0,1,\cdots,p$) and
$X^i$ (for $i=p+1,\cdots,9$). In the static gauge, $X^\mu$ are
identified with the worldvolume coordinates $\xi^\al$, $X^\mu=\xi^\mu$
and $X^i=constant$. Then from \rf{adX} it follows that in this gauge
$a^\ha_\mu =0$, so that the local frame orthogonal to the brane is
entirely spanned by $a^\ha_i\,$. Also from \rf{metrics} it follows
that $a^i_\ha a_j^\ha=\delta^i_j$. Hence the boundary conditions
\rf{DpBD}-\rf{DpFN} reduce to,
\beq
\ba{rrcl} 
{\rm Bosons:}\quad & \DX^i &=& 
\p_\tau X^i\br =0 \,, \\[.3cm] 
 &\NX_\mu &=& \big( E_{\mu M}\p_-X^M - E^T_{\mu M} \p_+ X^M 
\,-i\,\psi_-^M \p_M E_{\mu N}\psi^N_-      \\[.3cm]  
&&&\qquad +i\,\psi_+^M\p_M E^T_{\mu N}\psi^N_+ -i\,\eta \psi_+^M\p_\mu 
E_{MN}\psi_-^N \big)\br = 0  \,,       \\[.3cm] 
{\rm Fermions:} \quad&\Dpsi^i &=& \left(\psi_-^i +
\eta\psi_+^i\right)\br=0\,,\\[.3cm]  
 &\Npsi_\mu &=& \left(E_{\mu N}\psi_-^N -\eta E^T_{\mu N}\psi_+^N
                     \right)\br = 0 \,.
\ea
\label{DpAllsg}
\eeq

\subsection{Consistency with T-duality}  \label{ssBCtd}

T-duality has played an important role in the discovery and
understanding of D-branes
\cite{Dai:ua,Leigh:jq,Polchinski:1996na,Alvarez:1996up,Bergshoeff:1996cy,
Myers:1999ps}.  This is because, in constant backgrounds, T-duality
transformations are known to interchange D and N boundary conditions
thus naturally giving rise to Dp-branes in open string theory. This is
also the case for non-constant backgrounds in the absence of
fermions. Naturally, one expects the same to hold in the more general
case of N=1 supersymmetric worldsheet theory in X-dependent
backgrounds. This issue was first addressed in
\cite{Alvarez:1996up,Borlaf:1996na}, however the problem has remained
unresolved and subsequent work has not shed more light on
it. Considering the otherwise close connection between D-branes and
T-duality, this may be regarded as an unsatisfactory
situation. Besides, T-duality has been used to determine non-Abelian
scalar couplings in the worldvolume theory
\cite{Taylor:1999pr,Myers:1999ps}. As we will see later, the behaviour
of worldsheet boundary couplings under T-duality leads to an
understanding of these couplings. Motivated by all this, we now
show that the correct boundary conditions obtained in the previous
section are indeed consistent with T-duality in a transparent
way\footnote{\label{f2}The consistency of T-duality with the boundary
conditions has also been shown by the authors in \cite{ALZ}.} and also
write down the transformation of the boundary vectors. The results
will be used in the following section to construct the open string
boundary couplings to worldvolume fields.

Let us start with a D9-brane in background fields $\wt G_{MN}$ and
$\wt B_{MN}$ which do not depend on the $d=9-p$ coordinates $\wt X^i$,
but may depend on the remaining $p+1$ coordinates $\wt X^\mu$. The
Neumann boundary conditions are \rf{D9BN},\rf{D9FN},
\bea
&\wt E_{LM}\p_-\wt X^M \!\!  - \!\! \wt E^T_{LM} \p_+ \wt X^M \!\! 
-i\wt\psi_-^\mu \p_\mu \wt E_{LN}\wt\psi^N_- +i\wt\psi_+^\mu\p_\mu 
\wt E^T_{LN}\wt\psi^N_+ -i\eta \wt\psi_+^M\delta^\lambda_L\p_\lambda 
\wt E_{MN}\wt\psi_-^N\br=0\,, & \nonumber\\[.5cm] 
&\wt E_{LN}\wt\psi_-^N - \eta \wt E^T_{LN}\wt \psi_+^N \br = 0\,.&
\label{D9Nt}
\eea
We now perform T-duality transformations along the $d=9-p$ directions
$\wt X^i$. This should lead to the correct static gauge boundary
conditions \rf{DpAllsg} for Dp-branes in the T-dual background
$G_{MN}$, $B_{MN}$ with $X^i$ as the transverse directions. 

The effect of the dualities on the boundary vectors can be studied
in a systematic way in terms of the matrices $Q^M_{\mp N}$ and 
$P_{\mp MN}$ defined as,
\bea
Q^M_{-N} = \left(\ba{cc} \delta^{ik} E_{kj} &\delta^{ik}E_{k\nu} \\ 
         0^\mu_{\;j} & \delta^\mu_{\;\nu}\ea \right)\,,\,\,&\qquad&      
Q^M_{+N}=\left(\ba{cc}-\delta^{ik}E^T_{kj} &-\delta^{ik}E^T_{k\nu}\\ 
              0^\mu_{\;j} & \delta^\mu_{\;\nu} \ea \right)\,,\,\,\,\,
\label{Qpm} \\[.2cm]
P_{-MN}=\left(\ba{cc}  \delta_{ij} & 0_{i \nu} \\ 
         E_{\mu j}  & E_{\mu\nu} \ea \right)\,,\;\;\;\;&\qquad& \,\,\,\,     
P_{+MN}=\left(\ba{cc} -\delta_{ij} & 0_{i \nu} \\ 
                     E^T_{\mu j} & E^T_{\mu\nu} \ea \right)\,,      
\label{Ppm}
\eea
where, $E=G+B$ and $E^T=G-B$. It can be shown that $Q_\pm^{-1}$ are
given by the same expressions as for $Q_\pm$ but with $E$ replaced by
$\wt E$. The T-dual quantities are then related by
\cite{Hassan:mq,Hassan:1995je},  
\bea
&& \wt E = P_- Q^{-1}_- \,,\,\,\,\,\qquad \wt E^T = P_+ Q^{-1}_+ \,,
\label{EPQ}\\[.2cm]
&& \wt\psi^N_\pm = Q^N_{\pm L} \psi^L_\pm\,, \qquad
\p_\pm \wt X^N = Q^N_{\pm L} \p_\pm X^L -i \psi^\la_\pm \p_\la
Q^N_{\pm L} \psi^L_\pm \,.
\label{Qpsi}
\eea
It is easy to check that the transformation of $\p_\pm X$ follows from
that of $\psi_\pm$ under worldsheet supersymmetry transformations. 

Note the presence of the flat metric $\delta^{ik}$ in $Q_\pm$ (see for
example, \cite{Sen:1991zi, Hassan:1999mm}). Often this is not
explicitly shown when writing the T-duality transformations. But since
it will play a role later, we will briefly describe its origin:
T-duality transformations along coordinates $X^i$ commute only with an
$O(d)$ subgroup of the general coordinate transformations involving
$X^i$. Hence these $O(d)$ transformations of the original background
are identified with those of its dual and T-duality should explicitly
preserve this identification at all stages of the manipulation. To
make this manifest, the T-duality transformation formulae contain the 
$O(d)$ invariant metric to raise or lower indices.

Equipped with the above transformation rules, it is easy to show that
the D9-brane boundary conditions \rf{D9Nt} lead to the correct
Dp-brane boundary conditions under T-duality. For this, one can first
verify that
\beq
\wt\psi_+^M\,\p_\la\wt E_{MN}\,\wt\psi_-^N =
\psi_+^M\,\p_\la E_{MN}\,\psi_-^N \,.
\eeq
Then, using \rf{EPQ} and \rf{Qpsi} in \rf{D9Nt} one obtains, 
\beq 
\ba{l} 
P_{-LM}\p_-X^M-P_{+LM}\p_+ X^M \hfill \\ [.2cm]
\qquad\qquad\qquad
- i\psi_-^\la\p_\la P_{-LN}\psi^N_- + i\psi_+^\la\p_\la P_{+LN}\psi^N_+ 
-i\eta \psi_+^M\delta^\la_L \p_\la E_{MN}\psi_-^N \br = 0 \,,\\[.3cm]
P_{-LN}\psi_-^N - \eta P_{+LN}\psi_+^N \br = 0 \,.
\ea
\label{D9FN-Td}
\eeq
Now restricting to $L= i$ and $L=\mu$, respectively, and using the
form of the matrices $P_\pm$ in \rf{Ppm}, one recovers the correct
static gauge Dp-brane boundary conditions \rf{DpAllsg}. This shows the 
consistency of boundary conditions with T-duality.

In fact, one can go beyond boundary conditions and with equal ease 
obtain the T-duality action on the boundary vectors \rf{B}-\rf{Dpsi}
in static gauge. For backgrounds independent of $X^i$ and on using
\rf{Qpm}-\rf{Qpsi}, it is straightforward to show that the boundary
vectors transform as, 
\beq
\ba{llll}
\tilde\NX_\mu=\NX_\mu\,,  & \tilde\NX_i=2\delta_{ij}\DX^j\,,&
\tilde\Npsi_\mu=\Npsi_\mu \,, &
\tilde\Npsi_i=\delta_{ij}\Dpsi^j\,, \\[.3cm]
\tilde\DX^\mu=\DX^\mu\,,&\tilde\DX^i=\frac{1}{2}\delta^{ij}\NX_j\,,&
\tilde\Dpsi^\mu=\Dpsi^\mu \,, &
\tilde\Dpsi^i=\delta^{ij}\Npsi_j\,. 
\ea
\label{BVtd}
\eeq 

\section{Worldsheet Boundary Couplings} \label{sWSBC}

So far we have only considered open strings in metric and
antisymmetric tensor field backgrounds. In this section we will
describe a formalism for coupling the string worldsheet boundary to
D-brane worldvolume scalars and gauge fields. To elucidate the idea,
we start with the bosonic case and then move on to the N=1
supersymmetric theory with Abelian and non-Abelian worldvolume fields.
The structure of the boundary couplings and their behaviour under
T-duality points to an enlarged symmetry of the non-Abelian D-brane
worldvolume theories allowing us to obtain a covariant description of
the scalar field couplings. In this section, the worldvolume fields
are regarded as perturbations so the boundary conditions obtained
earlier remain unchanged. This allows us to deal with non-Abelian
worldvolume fields and obtain the associated open string vertex
operators.

\subsection{Worldsheet boundary couplings in bosonic theory} 
\label{ssWSBC-B}

Let us consider the coupling of the bosonic open string to the
worldvolume gauge fields $A_M$ and transverse scalars $\phi^M$.  
In flat space, $G_{MN}=\eta_{MN}$, $B_{MN}=0$, and in the static
gauge, worldvolume fields couple to the worldsheet boundary through,  
$$
\int_{\p\Si} d\tau A_\mu \p_\tau X^\mu - \int_{\p\Si} d\tau \phi_i
\p_\sigma X^i \,. 
$$ 
The scalar coupling follows from a vertex operator consideration
\cite{Dai:ua,Leigh:jq} as well as from T-duality
\cite{Polchinski:1996na}. This expression can be easily generalized to
a covariant one valid for curved D-branes and, for many purposes, is
adequate even in the presence of non-trivial backgrounds. However,
generalizations are needed when $B_{\mu i}\neq 0$. 
To discover the general form of the couplings, let us first understand
the flat space case from a different point of view: $\p_\tau X^M\br$
and $\p_\sigma X^M\br$ are two space-time vectors on the boundary of
the worldsheet. For a Dp-brane in flat background and in static gauge,
the Neumann and Dirichlet boundary conditions, $\p_\sigma X^\mu\br=0$
and $\p_\tau X^i\br=0$, project out some components of these vectors.
The surviving components $\p_\sigma X^i\br$ and $\p_\tau X^\mu\br$ are
precisely the operators on the worldsheet boundary to which the
D-brane scalars and vectors couple.

Having understood the flat space boundary couplings in this way,
it is straightforward to write the couplings in any general
background. In the bosonic case, the open string boundary conditions
involve the following two vectors on the worldsheet boundary: 
\beq
\N_L = E_{LN}\p_- X^N - E^T_{LN} \p_+ X^N \br \,,\qquad
\D^L = \p_\tau X^L \br \,.
\eeq
The N and D boundary conditions set to zero components of $\N$ and
$\D$ projected, respectively, along and normal to the D-brane
worldvolume: $\p_\al X^L \N_L= 0$, $a^\ha_L\D^L=0$. The surviving
components are $\N_\ha=a_\ha^L \N_L$ and $\D^\al=g^{\al\be}\p_\be X^M
G_{ML}\D^L$ to which the worldvolume fields should couple as 
$\N_\ha\Phi^\ha$ and $\D^\al\mA_\al$. In terms of the space-time
components $A_M$ and $\phi^M$ of the worldvolume fields given by 
\rf{wv-st}, the boundary couplings are  
\bea
S^{Dp}_{\p\Sigma} &=& \int d\tau\left[ A_M \D^M
+\frac{1}{2}\phi^M \N_M \right] \nonumber\\
&=&\int d\tau\left[ A_M \p_\tau X^M 
-\phi^L \left(G_{LM}\p_\sigma X^M - B_{LM}\p_\tau X^M\right)\right]\,.
\label{bcb}
\eea
The scalar field vertex operator now also has a $\p_\tau X^M$
contribution when the B-field has indices both along and transverse to
the brane, $a^M_\ha B_{MN}\p_\al X^N\neq 0$, which can be combined
with the gauge field part, 
\beq
S^{Dp}_{\p\Sigma} = \int d\tau\left[\left(A_M + \phi^L
  B_{LM} \right)\p_\tau X^M - \phi_N\p_\sigma X^N\right]\,. 
\label{bcb2}
\eeq
Later we will see that $A_M$ always appears in this combination, also
in the presence of supersymmetry and non-Abelian interactions. It may
seem appealing to get rid of the extra term by absorbing it in a
redefinition of $A_M$. But that would mean that non-trivial gauge
fields could be created simply by switching on transverse scalar
fields which should not be the case. 

To insure invariance of the action under $B_{MN}$ 2-form gauge
transformations, the transformation of $A_M$ should be modified such
that, 
\beq
\delta B_{MN}=\p_M\Lambda_N-\p_N\Lambda_M\,,\qquad \delta A_M 
=-\Lambda_M-\phi^L\left(\p_L\Lambda_M-\p_M\Lambda_L\right)\,.  
\label{2formGT}
\eeq
The extra term in $\delta A_M$ becomes relevant only when the 2-form
gauge transformation is not entirely restricted to the worldvolume. 
The origin of the modification will be explained below.

The boundary interaction \rf{bcb} is consistent with (and in fact
required by) the interpretation of $\phi^M$ as an infinitesimal shift
$X^M\rightarrow X^M+\phi^M$ in the position of the brane. Indeed, the
$\phi$-dependent part $S^\phi_{\p\Si}$ of the action can be generated
from the background part by shifting the coordinates to $X^M+\phi^M$
and retaining terms linear in the shift, 
$$
S_\Si[X+\phi] + S^A_{\p\Si} [X+\phi]\sim S_\Si[X]+ S^\phi_{\p\Si}[X]+ 
S^A_{\p\Si} [X]  
$$
Here, $S_\Si$ is the bosonic part of the worldsheet bulk action
\rf{actionCom}, and its variation under the shift gives rise to
$S^\phi_{\p\Si}$ along with a bulk term that vanishes by virtue of the
equation of motion. Hence adding $\frac{1}{2}\phi^M\N_M$ to the
boundary action is equivalent to the infinitesimal coordinate shift
$X^M \rightarrow X^M + \phi^M$. As for the gauge field part
$S^A_{\p\Si}$, to linear order in the worldvolume fields $S^A_{\p\Si}
[X+\phi]\sim S^A_{\p\Si} [X]$ and it could have been dropped. However,
its inclusion above clarifies the origin of the modification in the
transformation of $A_M$ in \rf{2formGT}: The left hand side of the
above equation is invariant under the usual NS-NS 2-form gauge
transformation with parameter $\Lambda(X+\phi)$. For the right hand
side, this implies the modified transformation \rf{2formGT}.

So far we have implicitly assumed that the world volume fields are
Abelian but the considerations can be generalized to the non-Abelian
case. When $A_M$ and $\phi^M$ are non-Abelian, then the boundary
action $S^{Dp}_{\p\Sigma}$ cannot be simply added to the bulk
worldsheet action, but should be inserted in the path integral through
a path-ordered Wilson line,
\beq
{\rm tr}\, {\cal P}\exp (i\,S^{Dp}_{\p\Sigma})
\label{wlb}
\eeq
The boundary action is still given by \rf{bcb} and the discussion
above, as well as the T-duality derivation to be described in the next
subsection continue to hold. If we still want to interpret this as
arising from the infinitesimal shift $X^M\rightarrow X^M+\phi^M$ (with
a non-Abelian $\phi^M$) in the bulk worldsheet action, then the above
path ordering along $\tau$ should be applied to the full action when 
regarded as a function of the shifted coordinate. The non-Abelian case 
will be discussed in more detail in the supersymmetric theory.

For later reference, we express the above boundary couplings in the
static gauge $X^\mu=\xi^\mu$, $X^i=const$. Then, as we have seen,
$\phi_i=\Phi^\ha a_{\ha i}$ and $\phi_\mu=\Phi^\ha a_{\ha\mu}=0$,
and the boundary action \rf{bcb2} becomes,
\bea
S^{Dp}_{\p\Sigma}
&=&\int d\tau\left[\left( A_\mu+\phi_i G^{iL}B_{L\mu}\right)
  \p_\tau X^\mu - \phi_i \p_\sigma X^i \right]
\label{bcbSG}
\eea
In this gauge the boundary conditions are $\N_\mu=0$, $\D^i=0$. 

\subsection{T-duality and restoration of covariance} \label{ssBBCtd}

We can go a step further and derive the scalar couplings of the last
section using T-duality. An offshoot of this is to clarify the
relation between the expression that results from T-duality and the
corresponding covariant expression (or its static gauge form). While
this relation is derived on the worldsheet boundary, it also holds in
the D-brane worldvolume theory where it enables us to promote
expressions obtained by T-duality to covariant ones. These
considerations are superfluous in the Abelian theory where they
simply follow from general coordinate transformation (GCT) invariance.
However, they have non-trivial consequences in the non-Abelian case
where the scalar couplings obtained by T-duality are seemingly
inconsistent with general covariance. For simplicity, we start with
the Abelian theory to explain the ideas and gradually generalize to
the N=1 non-Abelian case in the following subsections.

Let us start with a D9-brane with the boundary coupling,  
$$
S^{D9}_{\p\Sigma} = \int d\tau \wt A_M\p_\tau \wt X^M =
\frac{1}{2} \int d\tau \wt A_M\left(\p_+\wt X^M +\p_-\wt X^M\right)\,. 
$$ 
On general grounds, this is T-dual to the boundary coupling on a
Dp-brane with $p<9$ . To obtain the scalar field vertex operator on
the lower dimensional brane, we regard the gauge field as a
perturbation\footnote{\label{f3} If $\wt A_M$ is regarded as a large
  background field, it modifies the boundary conditions. Then
  following \cite{Alvarez:1996up}, T-duality results in a non-flat
  brane $X^\mu=\xi^\mu$, $X^i=\bar\phi^i(\xi)$ but without the
  boundary scalar couplings. For small $\bar\phi^i$, expanding around
  $X^i$ generates these coupling.} so that T-duality is given purely
in terms of the closed string backgrounds $G$ and $B$. Then, using the 
T-duality relation $\p_\pm\wt X^M =Q^M_{\pm N}\p_\pm X^N$ of the
bosonic theory with $Q_\pm$ given in \rf{Qpm} and making the usual  
identifications, $A_\mu=\wt A_\mu$, $\phi^i=\wt A_j\delta^{ji}$, one
obtains, 
\bea
S^{Dp}_{\p\Sigma} &=& \int d\tau\left[A_\mu \p_\tau X^\mu 
- \phi^i \left(G_{iM}\p_\sigma X^M-B_{iM}\p_\tau
X^M\right) \right]\nonumber \\
&=& \int d\tau\left[ A_\mu \D^\mu 
+\frac{1}{2} \phi^i \N_i \right] \,.
\label{bcbTD}
\eea 
On the one hand, from subsection \ref{ssBCtd} we know that the above
T-duality results in a Dp-brane in the static gauge in which both
$\phi^i$ and $\phi^\mu$ are non-zero \rf{aphisg}. On the other hand,
the expression in \rf{bcbTD}, which is a direct outcome of T-duality,
does not contain the components $\phi^\mu$ of the scalar field.
Indeed, although very similar to the static gauge expression
\rf{bcbSG} (which is obtained from the covariant action \rf{bcb2} on
gauge fixing the GCT), they are not yet exactly the same due to these
missing components of $\phi$. The relation between the two is relevant
to understanding the covariance of non-Abelian D-brane worldvolume
actions and will be spelt out below: 

To recover the static gauge action \rf{bcbSG} from the outcome of
T-duality, we have to lower the index on $\phi$ in \rf{bcbTD} to
$\phi_i = G_{iM}\phi^M$. For this one needs the missing scalar field
components $\phi^\mu$. Now, under T-duality the D9-brane boundary
condition $\wt\N_M=0$ goes over to static gauge Dp-brane boundary
conditions $\D^i=0$, $\N_\mu=0$. So we can add $\frac{1}{2}
\phi^\mu\N_\mu$ to the action without affecting its content. This
supplies the missing terms and insures that the outcome of T-duality
is consistent with the covariant action \rf{bcb2} or its gauge fixed
version \rf{bcbSG}. We have also seen that, for small $\phi^\mu$, the
addition of $\frac{1}{2} \phi^\mu\N_\mu$ to the boundary action is
equivalent to shifting $X^\mu$ to $X^\mu+\phi^\mu$. Hence the outcome
of T-duality is related to the covariant expression (or its static
gauge version) by a coordinate shift. Conversely, consider a Dp-brane
described by the embedding $X^M(\xi)$ with Abelian scalar fields
$\phi^M(\xi)$. This description is coveriant under GCT. If we fix the
static gauge right away, we end up with \rf{bcbSG}. However, we can
also use the GCT invariance to first make a transformation
$X^{'\mu}=X^\mu+\phi^\mu$ to eliminate $\phi^\mu$ and then go to the
static gauge. Now, the resulting expression coincides with the outcome
of T-duality \rf{bcbTD}.

To summarize, we saw that the outcome of T-duality can be written in a
GCT covariant form by adding to the boundary action terms that vanish
by the boundary conditions. For Abelian $\phi^\mu$ this is equivalent
to an ordinary coordinate transformation $X^\mu \rightarrow
X^\mu+\phi^\mu$ which can also be carried out in the corresponding
D-brane worldvolume theory to restore its manifest covariance. Since
these are known symmetries, the identification of their worldsheet
origin is redundant. However, when $\phi$ becomes a matrix in the
non-Abelian case, the shift is no longer an ordinary coordinate
transformation and its admissibility is not {\it a priori} evident in
the worldvolume theory. In this case, the worldsheet origin of the
transformation becomes crucial to insure its existence on the
worldvolume. This will be made more precise in the following
subsections.

\subsection{N=1 Supersymmetric Abelian worldsheet boundary couplings} 
\label{ssSupAb}

In the supersymmetric case, we start from the D9-brane boundary
couplings and obtain the rest by T-duality. From this a covariant
expression can be guessed leading to the N=1 boundary couplings in
general backgrounds and for any embedding. Alternatively these can be
obtained by supersymmetrizing the bosonic result. We will then discuss
the interpretation of scalars as coordinate shifts which is now more
subtle.

The N=1 supersymmetric D9-brane boundary coupling is given by
\cite{Callan:1988wz} 
\bea
S^{D9}_{\p\Sigma}&=&\int d\tau \left\{ A_M\p_\tau X^M -\frac{i}{4}
\left(\psi^M_- + \eta \psi^M_+ \right) F_{MN}\left(\psi^N_- + \eta
\psi^N_+ \right) \right\} \nonumber\\
&=& \int d\tau \left\{A_M \DX^M -\frac{i}{4}
\Dpsi^M  F_{MN} \Dpsi^N \right\} \,.
\label{D9bInt}
\eea 
Here we consider Abelian gauge fields postponing the non-Abelian case
to the next subsection. Under the NS-NS 2-form gauge transformation,
$\delta B=d\Lambda$, $\delta A=-\Lambda$, the variation of the first
term above is canceled by a boundary term that arises from the
variation of the bulk action \rf{actionCom}. Hence the term containing
$F_{MN}$ should be invariant by itself. Although this may not seem to
be the case at first sight, a closer examination shows that its
invariance can be made manifest. To see this, note that the D9-brane
Neumann boundary condition \rf{D9FN} implies
$$
\Dpsi^M\Npsi_M = \Dpsi^M B_{MN}\Dpsi^N-2\eta G_{MN}\psi_-^M\psi_+^N
=0\,. 
$$
Adding zero in this form to the boundary action one can write,
\beq
S^{D9}_{\p\Sigma}=\int d\tau\left\{A_M\p_\tau X^M -\frac{i}{4}
\Dpsi^M F_{MN}\Dpsi^N  -\frac{i}{4}\Dpsi^M\Npsi_M \right\}\,,
\label{D9bIntGN}
\eeq
without changing the dynamics. This contains the combination
$F_{MN}+B_{MN}$ which is manifestly invariant under 2-form gauge
transformations. In other words, the non-invariance of $F_{MN}$
conspires with that of the Neumann boundary condition \rf{D9FN} to
produce a gauge invariant result\footnote{If the worldvolume fields
are promoted to background fields, then they will enter the boundary
conditions rendering them invariant under 2-form gauge
transformations. However, as stated before, we regard $A$ and $\phi$
as perturbations to obtain their vertex operators in closed string
backgrounds and to avoid problems in the non-Abelian case.}. We will
see below that, besides making the 2-form gauge invariance manifest,
the added term is also needed to understand the nature of N=1
worldsheet couplings to D-brane scalars.

To obtain the N=1 supersymmetric boundary couplings on Dp-branes, we
apply T-duality to \rf{D9bInt} along the $d$ directions $X^i$.
Then, restricting to $\p_iA_M=0$, and using \rf{BVtd} with $A_\mu=\wt 
A_\mu$, $\phi^i=\wt A_j\delta^{ji}$, one gets,
\beq
\int_{\p\Sigma} d\tau \left\{A_\mu\DX^\mu
-\frac{i}{4}\Dpsi^\mu F_{\mu\nu}\Dpsi^\nu + \frac{1}{2}\phi^i
\NX_i -\frac{i}{2}\Dpsi^\mu \p_\mu \phi^i \Npsi_i \right\} \,.
\label{DpsTd}
\eeq
It is now easy to write the general covariant form of the couplings.
This is achieved, as in the bosonic case described in the last
subsection, by using the boundary conditions $\DX^i=\Dpsi^i=0$ and 
$\NX_\mu=\Npsi_\mu=0$ to complete the above action to,
\beq
S^{Dp}_{\p\Sigma}=\int_{\p\Sigma} d\tau \left\{A_M\DX^M
-\frac{i}{4}\Dpsi^M F_{MN}\Dpsi^N + \frac{1}{2}\phi^M
\NX_M -\frac{i}{2} \Dpsi^M \p_M \phi^N \Npsi_N \right\}\,.
\label{DpBAnc}
\eeq 
This expression is valid beyond T-duality and one can verify, using
\rf{BVSUSY}, that it is invariant under N=1 worldsheet supersymmetry.
Therefore it is the correct supersymmetric completion of \rf{bcb} and   
applies to any D-brane embedding provided the boundary vectors satisfy
the appropriate Dp-brane boundary conditions \rf{BVBC}. It is
understood that $\phi^M$ and $A_M$ are respectively normal and tangent
to the D-brane \rf{wv-st}, so that $\phi_M\p_\al X^M=0$ and
$A_Ma^M_\ha =0$. Boundary conditions imply that all derivatives acting
on $A_M$ and $\phi^M$ are of the form $\Dpsi^L\p_L = \Dpsi^\al\p_\al$
consistent with the fact that they live on the worldvolume and are
functions of $\xi^\al$. The covariance of the couplings will be
discussed later. 

As in the case of D9-brane, the term containing $F_{MN}$ is not
manifestly invariant under NS 2-form gauge transformations. To make
the invariance manifest, to this action we add   
\beq
S^{Dp,0}_{\p\Sigma}=\int_{\p\Sigma} d\tau \left\{
-\,\frac{i}{4}\Dpsi^\al\Npsi_\al \right\}\,,
\label{extra}
\eeq
which is zero by virtue of the boundary conditions and does not modify
the theory. The fact that this is the bare minimum required by 2-form
gauge invariance is important to get the correct scalar couplings and
hence is emphasized here. For example, $\Dpsi^\ha\Npsi_\ha$ also
vanishes but is not needed and so should not be included
\footnote{Note that T-duality gives $\tilde\Dpsi^M \tilde\Npsi_M =
\Dpsi^\mu \Npsi_\mu + \Npsi_i \Dpsi^i$ both of which vanish separately
by Dp-brane boundary conditions. Of these only the first term is
needed and is retained. Also, since the added term vanishes, so does
its supersymmetry variation. Hence we do not need to add its
supersymmetric completion $X^M\NX_M$ (for D9-brane). Thus we retain
the minimum required to make the 2-form gauge invariance manifest.}.

The boundary action \rf{DpBAnc} is manifestly supersymmetric although
in this form its covariance is not manifest. This is evident from the
presence of $\p_M\phi^N$ (instead of $\nabla_M\phi^N$) as well as
$\NX_M$ which, as noted earlier, does not transform as a vector. Also
the invariance of the $\phi$-dependent terms under the 2-form gauge
transformation is not manifest. To make both these symmetries
manifest, we note that after some manipulations equations \rf{B} and
\rf{F} can be written as,
\bea
\Npsi_M &=& G_{MN}\left(\psi^N_- -\eta\psi^N_+\right)+B_{MN}\Dpsi^N\,,
\nonumber\\[.1cm]
\NX_M &=& 2\left(B_{MN}\p_\tau X^N - G_{MN}\p_\sigma X^N\right) 
-i\,\Dpsi^N \Gamma^L_{NM} G_{LK}\left(\psi^K_- -\eta\psi^K_+\right)
\nonumber\\[.1cm] &&\qquad \qquad \qquad \qquad \qquad 
-i\,\Dpsi^N \p_N B_{ML}\Dpsi^L-i\eta\psi^N_+\psi^L_-H_{NLM}\,.
\nonumber 
\eea
Substituting in \rf{DpBAnc}, one recovers a manifestly covariant
expression for the action as, 
\bea
\ds S^{Dp}_{\p\Sigma}&=& \ds \int_{\p\Sigma} d\tau \Big\{
A_M^{(\phi)} \p_\tau X^M  - \frac{i}{4}\Dpsi^M F^{(\phi)}_{MN} \Dpsi^N   
\nonumber\\[.1cm]
&&\ds\qquad -\phi_N\p_\sigma X^N-\frac{i}{2}\,\Dpsi^\al\nabla_\al
\phi_M \left(\psi_-^M-\eta\psi_+^M\right)-\frac{i}{2}\eta\phi^M
\psi^N_+\psi^L_-H_{MNL} \Big\}\,. 
\label{DpBCoup}
\eea
As in the bosonic case, the gauge field always appears through this
shifted combination, 
\beq
A_M^{(\phi)} = A_M + \phi^L B_{LM}\,,
\label{Aphi}
\eeq
and $F^{(\phi)}_{MN}$ is the corresponding Abelian field strength.
Another feature of this boundary action is that the covariant
derivative $\nabla_\al\phi_M =\p_\al\phi_M -\p_\al X^N
\Gamma_{NM}^L\phi_L$ is in terms of the ordinary Christoffel
connection, without a torsion contribution. In fact, the torsion term
in \rf{DpBCoup} cannot be absorbed in the covariant derivative term in
this form.
Thus, the boundary action \rf{DpBAnc} is invariant under both general
coordinate transformations and NS 2-form gauge transformations (after
the addition of \rf{extra} and with the gauge field transformation
given by \rf{2formGT}).

Let us now understand the scalar couplings as coordinate shifts. In
the bosonic theory the boundary couplings of $\phi^M$ arose from the
shift $X^M\rightarrow X^M+\phi^M(\xi)$ in the worldsheet bulk action,
{\it i.e.}, the part with only closed string backgrounds.  In the N=1
case, in order to reproduce the $\phi$ terms in \rf{DpBAnc}, one has
to include the new features of the theory. First, consistency with the
supersymmetry transformation $\dels X^M=-\e^-\Dpsi^M$ implies that the
shift in $X^M$ should be accompanied by a corresponding shift in
$\Dpsi^M = \psi^M_-+\eta \psi^M_+$,
\beq
\Delta X^M = \phi^M(\xi)\,,\qquad\qquad
\Delta \Dpsi^M = \Dpsi^\al \p_\al\phi^M(\xi)\,.
\label{shift-abel}
\eeq 
To first order, the change $\Delta S$ of the worldsheet bulk action
\rf{actionCom} under these variations is given by \rf{actionBV}. This
is no longer zero since the variations $\Delta X^M$ and
$\Delta\Dpsi^M$ do not respect the boundary conditions. Besides, now
the action contains an extra purely background dependent piece
$S^{Dp,0}_{\p\Si}$ given by \rf{extra}. Although boundary conditions
set this term to zero, its variation $\Delta S^{Dp,0}_{\p\Si}$ under
the shifts is not zero (since we do not modify the boundary conditions
under the shift). The total variation of the purely background
dependent part of the action is then given by (for more details see
the discussion in the non-Abelian case below),
\beq
\Delta S + \Delta S^{Dp,0}_{\p\Si} =
\frac{1}{2}\int d\tau \left\{\Delta X^L\NX_L  
-i \Delta\Dpsi^L \Npsi_L \right\}.
\label{dS-abelian}
\eeq
Substituting for $\Delta X^M$ and $\Delta\Dpsi^M$ from
\rf{shift-abel}, we recover the $\phi$-dependent terms of the boundary
action \rf{DpBAnc}. This shows that in the N=1 theory too scalar
boundary couplings emerge from infinitesimal coordinate shifts,
provided supersymmetry and the presence of the extra term \rf{extra}
are taken into account.

The discussion in subsection \ref{ssBBCtd} of the relation between
T-duality and covariance of the scalar couplings applies with
appropriate generalizations: T-duality of the boundary conditions
leads to a Dp-brane in the static gauge, $X^\al=\xi^\al$ and
$X^i=const$. In general both $\phi^i$ and $\phi^\mu$ are non-zero.
However, the scalar couplings in \rf{DpsTd} resulting from T-duality
do not contain $\phi^\mu$ and are related to the manifestly covariant
action \rf{DpBAnc} (or its static gauge form) by the addition of
$\frac{1}{2} \NX_\mu \phi^\mu$ and $-\frac{i}{2} \Dpsi^\mu
\p_\mu\phi^\nu \Npsi_\nu$, both of which vanish by virtue of the
static gauge boundary conditions. But the addition of
these terms correspond to shifting $X^\mu$ to $X^\mu+\phi^\mu$, along
with the appropriate shifts in $\psi^\mu_\pm$. In the world volume
theory, this coordinate shift is the transformation needed to convert
the outcome of T-duality to the static gauge action and the above
discussion clarifies its worldsheet origin. In the next subsection we 
will generalize this to the non-Abelian theory where is acquires
non-trivial consequences. 

Let us now rewrite the action \rf{DpBCoup} in a form that highlights
the geometry of the brane. This is not relevant to the rest of the
paper and can be skipped. Geometrically, $\phi^M$ is normal to the
brane but the covariant derivative $\nabla_\al\phi^M$ has components
both normal and tangent to the brane and decomposes as  
\beq
\nabla_\al\phi^M = (\widehat\nabla_\al\Phi^\ha) a^M_\ha
- (\Phi_\ha \Omega^\ha_{\al\be}) g^{\be\gamma}\p_\gamma X^M\,.
\label{nab-decomp}
\eeq
In the first term, $\widehat\nabla_\al\Phi^\ha=\p_\al
X^M(\nabla_M\phi^N)a^\ha_N$ is by definition the covariant derivative
on the normal bundle. Its expression can be easily read off from this
definition as,
$$
\widehat\nabla_\al\Phi^\ha=
\p_\al\Phi^\ha + \omega^\ha_{\al \hb} \Phi^\hb \,,\qquad{\rm where,}
\qquad \omega^\ha_{\al\hb}=a^\ha_N\p_\al a^N_\hb+\p_\al X^M\Gamma^N_{MK}
a^\ha_N a^K_\hb\,.
$$ 
The second term contains the second fundamental form of the embedding
which can also be read off from \rf{nab-decomp} as,  
$$
\Omega^M_{\al\be}=\p_\al\p_\be X^M+\Gamma^M_{NK}\p_\al X^N\p_\be X^K\,.
$$
It gives the deviation of $\Phi^M$ from the normal under parallel
transport by the Christoffel connection. In string theory one often
encounters the torsionful connections $\Gamma^{\pm N}_{MK}=
\Gamma^{N}_{MK} \mp\frac{1}{2} H^{N}_{\,\,MK}$. It can be easily
checked that adding torsion to the left hand side of \rf{nab-decomp}
leads to torsion contributions to the normal bundle connection and to
the second fundamental form,   
\beq
\widehat\nabla^\pm_\al\phi^\ha=\widehat\nabla_\al\phi^\ha \pm
\frac{1}{2} \;\phi^M H_{M\al}^{\quad\;\;\ha} \,,\qquad
\Omega^{\pm K}_{\al\be}= \Omega^K_{\al\be}\mp\frac{1}{2}
H^K_{\,\,\,\al\be}\,, 
\label{nbcdt}
\eeq
where, $H_{M\al}^{\quad\;\;\,\ha}=H_{MLK}\p_\al X^L a^{K\ha}$ and 
$H^K_{\,\,\,\,\al\be}=H^K_{\,\,\,MN}\p_\al X^M \p_\be X^N$. Substituting
\rf{nab-decomp} in \rf{DpBCoup} and using \rf{nbcdt} along with the
Dirichlet boundary condition $\Dpsi^\ha= \psi_-^\ha+\eta\psi_+^\ha=0$,
one gets an alternative expression for the boundary action as 
\bea
\ds S^{Dp}_{\p\Sigma}&=&\ds\int_{\p\Sigma} d\tau \Big\{
A_M^{(\phi)} \p_\tau X^M  - \frac{i}{4}\Dpsi^M F^{(\phi)}_{MN} \Dpsi^N   
-\phi_N\p_\sigma X^N+i\psi^\al_+\widehat\nabla^+_\al\Phi^\ha \psi_{+ \ha}
\nonumber\\
&&\ds \qquad -i\psi^\al_- \widehat\nabla^-_\al\Phi^\ha \psi_{- \ha}  
+\frac{i}{2}\eta \psi^\al_+\psi^\be_-\Omega^{+M}_{\al\be}\phi_M +
\frac{i}{4}(\psi^\ha_+\psi^\hb_+ +\psi^\ha_-\psi^\hb_-)
H_{\ha\hb M}\phi^M \Big\} \,.
\label{Dp-alt}
\eea 
As usual, all indices are raised and lowered using the corresponding
metrics $G_{MN}$, $g_{\al\be}$ and $\delta_{\ha\hb}$. 

\subsection{N=1 Supersymmetric non-Abelian boundary couplings}   
\label{ssSupNA}

The more interesting case from the point of view of D-brane
worldvolume theories is the boundary couplings of non-Abelian
worldvolume fields and their behaviour under T-duality. The procedure
for constructing the couplings is similar to the Abelian case and will
be carried out here. The interpretation of scalar couplings as
coordinate shifts is more subtle and will be addressed in the next
subsection. 

The starting point is the path-ordered supersymmetric Wilson line for
the D9-brane which is inserted in the path-integral measure
\cite{Callan:1988wz}, 
\beq
{\rm tr}\, {\cal P}\exp (i\,S^{D9}_{\p\Sigma})={\rm tr}\, {\cal P}\exp 
\left(i\,\int_{\p\Sigma} d\tau\left\{A_M\DX^M-
\frac{i}{4}\Dpsi^M F_{MN}\Dpsi^N\right\}\right)\,.
\label{wls}
\eeq 
Here, $A_M=A^a_M\lambda^a$ are non-Abelian gauge fields with
$\lambda^a$ denoting the gauge group generators and $F_{MN} = \p_M A_N  
-\p_N A_M +i[A_M\,, A_N]$. The invariance of $S^{D9}_{\p\Sigma}$ under
N=1 supersymmetry requires an unconventional transformation of
$\lambda^a$. The combined supersymmetry transformations are
\cite{Callan:1988wz},
\beq
\dels X^M = -\e^-\Dpsi^M\,,\quad \dels\Dpsi^M=-2i\e^-\p_\tau X^M\,,
\quad \dels\lambda^a=-i\e^-\Dpsi^L\,[A_L\,,\lambda^a]\,.
\label{stna}
\eeq

To obtain the non-Abelian Dp-brane boundary couplings from this, we
take $A_M$ to be independent of $d$ coordinates $X^i$ and apply
T-duality along these directions. Then, using \rf{BVtd} and making the
usual identifications $\wt A_\mu=A_\mu$, $\wt A_i=\delta_{ij}\phi^j$
(where the origin of $\delta_{ij}$ was explained in section
\ref{ssBCtd}), the action $S^{D9}_{\p\Sigma}$ in \rf{wls} yields, 
\bea
&&\int_{\p\Sigma} d\tau \Big\{A_\mu\DX^\mu
-\frac{i}{4}\Dpsi^\mu F_{\mu\nu}\Dpsi^\nu \nonumber\\[.1cm]
&& \qquad\qquad 
+\frac{1}{2}\phi^i    \NX_i -\frac{i}{2}\Dpsi^\mu
\left(\p_\mu \phi^i+i\left[A_\mu\,,\phi^i\right]\right)\Npsi_i
+\frac{1}{4}\Npsi_i\left[\phi^i\,,\phi^j\right]\Npsi_j\Big\}\,.
\label{DpsTdna}
\eea
As before, the missing components can be inserted by using the
static gauge boundary conditions $\DX^i=\Dpsi^i=0$,
$\NX_\mu=\Npsi_\mu=0$, without affecting the content of the theory,
\bea
&&S^{Dp}_{\p\Sigma}=\int_{\p\Sigma} d\tau \Big\{A_M\DX^M
-\frac{i}{4}\Dpsi^M F_{MN}\Dpsi^N \nonumber\\[.1cm]
&& \qquad
+\frac{1}{2}\phi^M \NX_M -\frac{i}{2}\Dpsi^M\left(\p_M \phi^L
+ i\left[A_M\,,\phi^L\right] \right)\Npsi_L+\frac{1}{4}\Npsi_M
\left[\phi^M\,,\phi^N\right]\Npsi_N\Big\}\,.
\label{Dpsna}
\eea
This final expression can be reinterpreted as the general form of the
boundary couplings which goes beyond T-duality and is valid for any 
background and any embedding. As a check, one can verify its
invariance under N=1 worldsheet supersymmetry that acts on the
boundary vectors as in \rf{BVSUSY}, coupled with an unconventional 
transformation of the gauge group generators $\lambda^a$ that can be
surmised from the action of T-duality on the last equation in
\rf{stna},
\beq
\dels\lambda^a=-i\e^-\Dpsi^L\,[A_L\,,\lambda^a]
-i\e^-\Npsi_L\,[\phi^L\,,\lambda^a]\,.
\label{stnaDp}
\eeq

As in the Abelian case, to the action \rf{Dpsna} we should further add
the extra term \rf{extra} which vanishes by virtue of Dp-brane
boundary conditions, but which makes the NS 2-form gauge invariance of
the $F_{MN}$-term manifest.

After some manipulations very similar to the N=1 Abelian case, the
action \rf{Dpsna} can be written in a manifestly general coordinate
invariant form,
\bea
&\ds S^{Dp}_{\p\Sigma}= \ds \int_{\p\Sigma} d\tau \Big\{
A_M^{(\phi)} \p_\tau X^M  - \frac{i}{4}\Dpsi^M F^{(\phi)}_{MN} \Dpsi^N   
-\phi_N\p_\sigma X^N-
\frac{i}{2}\,\Dpsi^\al \nabla_\al^{(A^\phi)}
\phi_M \left(\psi_-^M-\eta\psi_+^M\right)&
\nonumber\\[.1cm]
&\ds\qquad \qquad \qquad \qquad \qquad \qquad 
+i\eta \psi^N_+\psi^L_- \left(\left[\phi_N,\phi_L\right]
-\frac{1}{2}\phi^M H_{MNL}\right)\Big\}\,. &
\label{DpCNA}
\eea
The covariant derivative is now,
\beq
\nabla_\al^{(A^\phi)}\phi_M=\p_\al X^L\left(\p_L\phi_M - \Gamma_{LM}^N\phi_N             
  +i \, [ A^{(\phi)}_L , \phi_M ] \right)\,,
\label{nbGA}
\eeq
and throughout the gauge field appears through the shifted
combination,    
$$
A_M^{(\phi)}= A_M + \phi^L B_{LM}\,, 
$$ 
including in the field strength, $F_{MN}^{(\phi)}=\p_M
A_N^{(\phi)}-\p_N A_M^{(\phi)} + i\, [A_M^{(\phi)},A_N^{(\phi)}]$. 
To get the $\left[\phi_N,\phi_L\right]$ term in this form, we have
used $\Dpsi^M\phi_M=0$. The action can also be rewritten in the form
\rf{Dp-alt}. 

\subsection{Non-Abelian scalar couplings as coordinate shifts}
\label{ssNaCX}

The open string boundary coupling to non-Abelian scalar fields 
also follow from a shift of the coordinates $X^M$ by the matrix
valued fields $\phi^M$, although with some subtleties. The implication 
is however more drastic as it leads to the notion of non-Abelian 
coordinate shifts as a symmetry of the D-brane worldvolume theory.
This in particular allows us to relate worldvolume scalar 
couplings obtained by T-duality to the corresponding covariant 
expressions. The shift is no longer a part of general coordinate
transformations.   

Consider the coordinate shift $\triangle X^M=\phi^M$ performed in the
purely background dependent part of the action. This is an unusual
shift as it involves adding a non-Abelian matrix $\phi^M
=\phi^{aM}\lambda^a$ to the ordinary coordinate $X^M$. To carry this
out, the path-ordering in the Wilson line \rf{wls} should be extended
to include the full worldsheet action factor $\exp(iS)$ in the path
integral. In practice, path ordering will not cause much complication  
since we are interested only in first order variations.  To
preserve supersymmetry, the shift in $X^M$ should be accompanied by a
shift $\triangle\Dpsi^M=\triangle(\psi^M_- + \eta\psi^M_+)$ of the
fermions. This is determined by extending the supersymmetry
transformation of $X^M$ to its variation, $\dels \triangle X^M =
-\e^-\triangle\Dpsi^M$. Using \rf{stnaDp}, the supersymmetry variation
of $\phi^M= \phi^{Ma} \lambda^a$ is
$$
\dels\phi^N=-\e^-\left(\Dpsi^L \p_L\phi^N + i\,\Dpsi^L [A_L,
\phi^N] + i\,\Npsi_L [\phi^L,\phi^N] \right)\,.
$$
From this one can read off $\triangle\Dpsi^M$, leading to the bosonic
and fermionic shifts, 
\beq \ba{c}
\triangle X^M=\phi^M\,, \\[.2cm] 
\triangle\Dpsi^M=\Dpsi^L\left(\p_L\phi^M+i\,[ A_L\,,\phi^M]
\right)  + i\,\Npsi_L [\phi^L\,,\phi^M]\,.
\ea
\label{nabshift}
\eeq

To first order in the shifts, the variation of the worldsheet bulk
action is given by \rf{actionBV},
\beq 
\triangle S =\frac{1}{2}\int_{\p\Sigma}d\tau \Big[ \triangle X^L \NX_L + 
\frac{i}{2}\,\Dpsi^L\triangle\Npsi_L -\frac{i}{2}\,\triangle\Dpsi^L 
\Npsi_L \Big]\,.
\label{dSbulk}
\eeq
This is non-zero since only the boundary vectors satisfy boundary
conditions but not the shifts $\triangle X$ and $\triangle\Dpsi$. To
this we have to add the variation of the purely background term
\rf{extra} required by NS 2-form gauge invariance. While this term
and part of its variation vanish due to the boundary condition
$\Npsi_\al=0$, the non-zero piece in the variation is given by,  
\bea
\triangle S^{Dp,0}_{\p\Sigma}&=&-\,\frac{i}{4}\int_{\p\Sigma}d\tau 
\left\{\Dpsi^\al \triangle\left(\p_\al X^L\Npsi_L\right) \right\}
\nonumber\\
&=&-\,\frac{i}{4}\int_{\p\Sigma}d\tau\left\{\Dpsi^L\triangle(\Npsi_L)
+ \Dpsi^\al \triangle\left(\p_\al X^L\right) \Npsi_L \right\}\,,
\label{dextra}
\eea
where we have used $\Dpsi^\al\p_\al X^L = \Dpsi^L$. Then, the expected 
scalar field boundary couplings $S^{(\phi)}=\triangle S+ \triangle
S^{Dp,0}_{\p\Sigma}$ are
\beq 
S^{(\phi)}=\frac{1}{2}\int_{\p\Sigma}d\tau \left\{\triangle X^L \NX_L  
-\frac{i}{2}\,\triangle\Dpsi^L \Npsi_L 
-\frac{i}{2}\Dpsi^\al\triangle\left(\p_\al X^L\right)\Npsi_L 
\right\}\,.
\label{S-phi}
\eeq
If the shift is Abelian, then $\triangle\left(\p_\al X^L\right)=
\p_\al(\triangle X^L)=\p_\al\phi^L$, as can be verified by noting that
the supersymmetry transformations \rf{BVSUSY} also hold for
$\triangle\Dpsi$ and $\triangle\DX$. In this case we end up with the
expression \rf{dS-abelian} which was used in the N=1 Abelian case. 

In the non-Abelian theory $\triangle$ and $\p_\al$ do not commute and
the above identification is not consistent with supersymmetry. One can
attempt to find the correct expression for $\triangle\left( \p_\al
X^L\right)$ by considering instead $\triangle\left(\p_\tau X\right)
\equiv\triangle\DX$ and taking the right hand side to be defined by
the supersymmetry transformation $\dels (\triangle\Dpsi) =-2i\e^-
\triangle\DX$. Then using \rf{nabshift} one finds that $\triangle\DX^L
= \p_\tau X^M (\p_M\phi^L +i \,[A_M,\phi^L]) +\cdots$. If we ignore
the extra terms this suggests,  
\footnote{\label{f6} This approach is not fully consistent since a
  straightforward generalization of \rf{BVSUSY} to non-Abelian
  quantities does not lead to a closed set of equations, although we
  can still get some information from it. The complete expression for 
  $\triangle\DX^L$ obtained in this way is:\\[.1cm] 
\centerline{
$\p_\tau X^M D_M\phi^L+\frac{i}{2}\NX_P[\phi^P,\phi^L]
+\frac{1}{4}\Dpsi^K\Dpsi^M[F_{KM},\phi^L] -\frac{1}{2}\Npsi_P\Dpsi^M
[D_M\phi^P,\phi^L] -\frac{i}{2}\Npsi_P\Npsi_K[\phi^K,[\phi^P,\phi^L]],
$}\\[.1cm]
where $D_M=\p_M +i[A_M,\,\,\,\,]$.} 
$$
\triangle(\p_\al X^L)= \p_\al X^M\left(\p_M\phi^L+i\,[A_M,\phi^L]
\right)\,.
$$ 
Substituting this along with \rf{nabshift} in \rf{S-phi} one recovers
the correct $\phi$-dependent terms in the non-Abelian boundary action.  
While this leads to the correct result, the above derivation is
not fully satisfactory in view of footnote \ref{f6}. One way of making
this rigorous would be to find a closed form of the supersymmetry
transformations \rf{BVSUSY} when applied to non-Abelian quantities.
We will not follow this approach but rather describe an alternative
derivation which, although formal, is more illuminating: 

The problem has its origin in the addition of $\phi^M$, which 
transform in the adjoint representation of the gauge group, to   
gauge singlets $X^M$. A way out would be to make $\phi^M$ behave more
like gauge singlets. This can be achieved at least formally, albeit at
the expense of locality. Consider a gauge transformation $U$ in the 
non-Abelian theory,
\bea
&\mA^\prime_\al=U^{-1}\mA_\al U - i\,U^{-1}\p_\al U\,, &\nonumber\\
&\phi^{\prime M} = U^{-1}\phi^M U \,.&
\label{AphiU}
\eea  
On the worldsheet boundary, $\mA_\al$ and $\phi$ depend on $\tau$
through $\xi^\al(\tau)$. We can always choose $U$ such that at a given
point $\tau$ the gauge field vanishes (although the field strength is
non-zero). This is the analogue of Riemann normal coordinates for the
gauge bundle. For this, consider a path $\xi(\bar\tau)$ from some
$\tau_0$ to $\tau$ and choose $U$ as the Wilson line,
\beq
U(\tau,\tau_0)={\cal P}\exp\,(-i\int_{\tau_0}^\tau\mA_\tau\,
 d\bar\tau)\,,
\label{U}
\eeq
where $\mA_\tau=\mA_\al\p_\tau\xi^\al=A_M\p_\tau X^M$. Keeping the 
ordering in mind, one can see from \rf{AphiU} that at the point
$\tau$, $\mA^\prime_\al(\tau)=0$ and furthermore, $\phi^{\prime
  M}(\tau)= U(\tau_0,\tau) \phi^M(\tau) U(\tau,\tau_0)$ is invariant
under local gauge transformations that are localized  in the
neighbourhood of $\tau$ but which vanish at $\tau_0$ (in particular
one can choose $\tau_0$ at infinity and ignore its  effect). Now the 
$\phi^{\prime M}$, although non-local, are singlets at $\tau$ and are
the natural objects by which the coordinates can be shifted. Thus in 
this gauge the shifts \rf{nabshift} become,  
\beq\ba{c}
\triangle X^M=\phi^{\prime M}\,,  \\[.2cm]
\triangle\Dpsi^M=\Dpsi^L \p_L\phi^{\prime M}
+i\, \Npsi_L \left[ \phi^{\prime L}\,,\phi^{\prime M}\right]\,.
\ea
\label{nabshiftp}
\eeq
Treating $\phi^\prime$ as a gauge singlet at $\tau$, we write
$\triangle\left(\p_\al X^L\right)=\p_\al(\triangle X^L) = \p_\al
\phi^{\prime L}$ as in the Abelian case. Then substitution in
\rf{S-phi} reproduces the boundary couplings \rf{Dpsna} to
$\phi^{\prime M}$ in the gauge where $A^\prime_M=0$. The gauge field
can be reinstated by the inverse gauge transformations $U^{-1}$ on
noting that,   
\beq
\p_\al\phi^{\prime M}=U^{-1}(\p_\al\phi^M+i\,[\mA_\al,\phi^M])U\,.
\label{dphip}
\eeq
To summarize, we have seen that the N=1 worldsheet boundary couplings
to non-Abelian worldvolume scalars are reproduced by shifting the
$X^M$ by the matrices $\phi^M$ in the background part of the action.
Carried out consistently, this also implies the shift $\p_\al X^M
\rightarrow \p_\al X^M + \p_\al \phi^M +i\,[\mA_\al,\phi^M]$. 

We can now discuss the issue of T-duality and covariance which is very
similar to the bosonic theory with Abelian fields, although now the
implications are non-trivial: T-duality on D9-branes gives rise to
Dp-brane boundary conditions in static gauge $X^\mu= \xi^\mu$,
$X^i=const$. While the scalar field components $\phi^i$ and $\phi^\mu$
are generically non-zero in this gauge (as summarized in \rf{aphisg}),
T-duality is capable of generating only $\phi^i$. In particular, it
produces the Dp-brane boundary action \rf{DpsTdna} in which the
$\phi^\mu$ do not appear. These missing components are needed if we
are to reproduce the static gauge version of the covariant action
\rf{DpCNA}, insuring consistency with general covariance.  We have
seen that the missing components $\phi^\mu$ can be included in the
action through the addition of terms like $\phi^\mu\NX_\mu$ that
vanish by virtue of the boundary conditions. This has no effect on the
physics and in this sense is a symmetry operation. Finally, we have
seen that the same operation can also be implemented by shifting the
coordinates by the matrix valued field $X^\mu\rightarrow X^\mu
+\phi^\mu$, which when performed correctly, also results in $\p_\al
X^\mu \rightarrow \p_\al X^\mu + \p_\al \phi^\mu
+i\,[\mA_\al,\phi^\mu]$.

D-brane worldvolume action is determined by the string worldsheet
theory and therefore shares its behaviour under T-duality and inherits
the above symmetry. Since worldsheet boundary conditions do not have a
direct counterpart in the worldvolume theory, it is the implementation
of the symmetry by the matrix-valued coordinate shift that provides
its worldvolume realization. Unlike on the worldsheet, the extra terms
generated by the shift on the worldvolume do not vanish identically,
indicating that the shift is part of a non-trivially realized group of
matrix-valued coordinate transformations, MCT. These generalize
ordinary general coordinate transformations, GCT.  The worldsheet
considerations above uncover only a part of the MCT, enough to address
the issue of covariance, but do not clarify its general group
structure. The implications of this will be discussed in the next
section.

\section{Covariant Coupling of Scalars in Worldvolume Theories}
\label{sDbrane} 

This section addresses the issue of the apparent incompatibility of
non-Abelian scalar couplings with general covariance which arises in
the known form of the worldvolume theory. As advertised, the puzzle is
resolved by the extra symmetry observed on the worldsheet boundary. We
first consider the Abelian case to demonstrate the method and then
turn to the non-Abelian theory. At the end, we present the general
picture which puts everything in context.

\subsection{Review of scalar couplings in Abelian worldvolume theory} 
\label{ssDbAb}

Here we review the structure of the Abelian D-brane worldvolume
theory with emphasis on the relation between T-duality and
covariance, in order to set the stage for the discussion of the
non-Abelian case in the next subsection. We will also comment on the
nature of the scalar field as a coordinate difference on the
worldvolume.  

The Abelian theory on the worldvolume of a single Dp-brane is given by
a sum of the Dirac-Born-Infeld (DBI) \cite{Leigh:jq} and
Chern-Simons (CS) actions \cite{Douglas:1995bn}
\beq
\int d^{p+1}\xi\, e^{-\varphi} \sqrt{det(P[E]+\mF)} + 
\int P[C^\prime]\wedge e^\mF\,, 
\label{abWV}
\eeq
where higher order curvature corrections have been ignored.
The structure of the action depends on the brane geometry given by the   
embedding $X^M(\xi^\al)$ through $P[\cdots]$ which denotes the
pull-back of space-time tensors to the brane worldvolume. For a
space-time tensor $V_{M_1\cdots M_n}$ (which stands for $E=G+B$ in the
DBI part and for the Ramond-Ramond potentials $C^{\prime(n)}$ in the
CS part), it denotes,
\beq
P[V]=\p_{\al_1} X^{M_1}\cdots\p_{\al_n}X^{M_n}\,V_{M_1\cdots M_n}\,.
\label{pb}
\eeq
For the RR potentials we use the notation, $C^\prime=C\wedge e^B$,
where $C$ are invariant under NS 2-form gauge transformations. $\mF =
d\mA$ is the gauge field strength on the worldvolume (see subsection 
\ref{ssCov} for conventions). In the Abelian theory, the coupling of
D-brane charges $e^\mF$ to background RR potentials is described by
exterior multiplication. This action is manifestly invariant under
space-time general coordinate transformations (GCT).  

Although the scalar fields do not explicitly appear in the above
description, string worldsheet considerations suggest that at least to
first order, they could be hidden as coordinate shifts in the
embedding functions $X^M(\xi)$ \cite{Leigh:jq}. To see this
explicitly, one starts with the D9-brane worldvolume action with
gauge fields $\tilde A_M$ and obtains the Dp-brane action by
T-dualizing along the coordinates $X^i$, leaving $X^\mu$ unchanged. 
After identifying $\tilde A_\mu=A_\mu$, $\tilde A_i= \delta_{ij}
\bar\phi^j$, a small generalization of the procedure in  
\cite{Alvarez:1996up,Bergshoeff:1996cy} yields the Dp-brane with the
scalar couplings (see footnote \ref{f3}). This is still given by
\rf{abWV} but now in the semi-static gauge, $X^\mu = X^\mu(\xi^\al)$,
$X^i = \bar\phi^i (\xi^\al)$ where the pull-backs take the form,
\beq 
P[V]_{T-dual}=
\cdots \p_\al X^\mu  V_{\mu \cdots} + 
\cdots \p_\al \bar\phi^i V_{i \cdots} \,.
\label{ssg}
\eeq  
Note that in the above, the scalar fields $\bar\phi^i$ naturally
appear as {\it coordinate differences} and hence are distinguished
from the scalar fields $\phi^M$ (or $\Phi^\ha$) of the worldsheet 
boundary theory which are {\it vectors} normal to the brane. The
relation between the two will be discussed later.

We now make the connection between covariance, T-duality and scalar
couplings even more explicit: \rf{ssg} gives the coupling of scalar
fields in a specific GCT gauge, one that followed from T-duality. One
may ask how the scalars couple in general? From worldsheet
considerations it is clear that on a Dp-brane defined by the embedding
$X^M(\xi)$ there will exist scalar fields $\phi^M$ (or $\bar\phi^M$,
to be specified below). Since both $X^M(\xi)$ and the scalars
correspond to brane shape and position, there is some freedom in how
much of this information one encodes in each one the two. It is then
evident that the general coupling of scalars in the Abelian
worldvolume action should be through $X^M+\bar\phi^M$ and the
expression for the pull-backs to be used in \rf{abWV} is
\beq
P[V]_{X+\bar\phi}=\p_{\al_1} (X^{M_1}+\bar\phi^{M_1})\cdots
\p_{\al_n}(X^{M_n}+\bar\phi^{M_n})\,V_{M_1\cdots M_n}(X+\bar\phi)\,. 
\label{pbs}
\eeq
This is a covariant expression for the coupling of scalars as long as
$\bar\phi^M$ are regarded as coordinate differences. The outcome of
T-duality \rf{ssg} is a gauge fixed version of the above for which we
have to fix the static gauge and also get rid of $\phi^\mu$ by the
coordinate transformation $X^\mu \rightarrow X^\mu-\phi^\mu$.
Conversely, the $\phi^\mu$ can be reinstated in the T-duality
expression \rf{ssg} by the coordinate shift $X^\mu \rightarrow X^\mu
+\phi^\mu$, partially undoing the gauge fixing. It is in this sense
that the Abelian Dp-brane action obtained by T-duality is compatible
with covariance.

The coordinate shift $\delta X^\mu=\phi^\mu$ that insured the
compatibility of T-duality and general covariance is a part of GCT,
which is a symmetry of the Abelian worldvolume action \rf{abWV}. In
section \ref{sWSBC} the same shift appeared as an invariance of the
worldsheet boundary action obtained by T-duality and served the same
purpose as on the worldvolume. The obvious conclusion is that even if
we did not know about the general covariance of the worldvolume
action, this property of the boundary theory would have been enough to
indicate the existence of such a shift symmetry on the worldvolume
enabling us to promote the outcome of T-duality in \rf{ssg} to a
covariant expression. This is an example of how our worldsheet
considerations can be applied to the worldvolume theory. It also shows
that the shift symmetry observed on the worldsheet is part of a larger
symmetry of the worldvolume theory, in this case the GCT. The
generalization to non-Abelian theory would imply the existence of a
group of matrix-valued coordinate transformations, MCT, only a part of
which is observed in the worldsheet theory.

We will now remark on the relation between the scalars $\phi^M$ that 
appear in the worldsheet boundary action and the $\bar\phi^M$ of the
worldvolume action. $\phi^M$ transforms as a vector under GCT as is
evident from the structure of the worldsheet operator to which it
coupled. On the other hand, $\bar\phi^M$ appears as a coordinate
difference and transforms accordingly. A manifestation of the
difference is that we encounter ordinary derivatives of $\bar\phi^M$
but only covariant derivatives of $\phi^M$. As suggested in
\cite{DeBoer:2001uk} it is natural to think of the relation in terms
of the Riemann normal coordinates \cite{petrov, eisenhart} which also
appears in other contexts in the physics literature   
\cite{Alvarez-Gaume:hn, Mukhi:1985vy}. This allows one to express a
coordinate difference $\Delta X^M$ in terms of vectors $u^M$
(that acquire the interpretation of tangent vectors at $X$ to
geodesics from $X$ to $X+\Delta X$), 
\beq
\Delta X^M=u^M - \sum_{n=2}^\infty \frac{1}{n!}\Gamma^M_{L_1\cdots
L_n} u^{L_1}\cdots u^{L_n}\,.
\label{rnc}
\eeq
Here, $\Gamma^M_{L_1\cdots L_n}=\nabla_{(L_1}\cdots\nabla_{L_{n-2}}
\Gamma^M_{L_{n-1}L_n)}$, with the covariant derivatives acting only on
the lower indices, are evaluated at a point $X^M(\xi)$ on the D-brane
and we regard $\Delta X$ as a displacement from this point. The
manipulations involving Riemann normal coordinates require that $u^M$
spans all directions, so we write $u^M=u^\al\p_\al X^M+\phi^M$. Only
at the end of the day can we restrict the results to the normal
directions, $u^M=\phi^M$, in which case $\Delta X^M =\bar\phi^M$
becomes a displacement away from the brane. Using the Riemann normal
coordinate formalism, one can see that \cite{Alvarez-Gaume:hn,
Mukhi:1985vy},
\beq
\p_\al\bar\phi^M=\nabla_\al\phi^M-\frac{1}{3}\p_\al X^K 
R_{L_1L_2K}^{M} \phi^{L_1}\phi^{L_2} + \cdots 
\label{dphib}
\eeq
where $\nabla_\al\phi^M=\p_\al\phi^M+\p_\al X^N\Gamma^M_{NL}\phi^L$,
and the ellipses denote terms with higher powers and derivative of the
curvature tensor. It is the right hand side of this equation that
should emerge from an appropriate worldsheet calculation of the
worldvolume action. Note that the covariant derivative above is the
same as the one in \rf{DpBCoup}; not restricted to the normal bundle
and without a torsion contribution. To lowest order $\phi$ and
$\bar\phi$ are the same and it was only to this order that the
worldsheet manipulations were carried out.

\subsection{Covariance of non-Abelian worldvolume theory}
\label{ssDbnA}

In this subsection we finally address the issue of covariance of the
scalar couplings in non-Abelian worldvolume theory. Geometrically, a
stack coincident D-branes is still described by the embedding
functions $X^M(\xi)$, and the non-Abelian worldvolume fields $A_M$ and
$\phi^M$ are, respectively, tangent and normal to the stack. The
covariance of this description under general coordinate
transformations (GCT) is reflected in the covariance of the worldsheet
boundary couplings \rf{DpCNA} which, in turn, determine the D-brane
worldvolume action (say, through a manifestly covariant background
field computation as in \cite{Alvarez-Gaume:hn,Leigh:jq}. The obvious
implication is that the worldvolume theory should at least exhibit
ordinary general coordinate invariance, besides other larger
non-Abelian symmetries that it may also possess.

In practice, the couplings of non-Abelian scalars in the worldvolume
action are obtained by T-duality or computations around flat
background
\cite{Taylor:1999gq,Taylor:1999pr,Myers:1999ps,Garousi:2000ea}, where 
covariance is not manifest. Even then, one normally expects the final
results to be consistent with general covariance in the sense that
they are obtainable from a GCT invariant action on going to a specific
coordinate gauge, as happens in the Abelian theory. This however is
not the case in the non-Abelian theory. To see this, let us consider
the couplings of non-Abelian scalars obtained in
\cite{Myers:1999ps}. Instead of writing the full action, we
concentrate on the general structure of the couplings involving the
scalars. These are determined in large part by T-duality and, in the
static gauge, where $X^\mu=\xi^\mu$ along the brane and $X^i=const$ in
the ``transverse'' directions, they appear as
\bea
&V_{\mu\,\cdots}+D_\mu\phi^i\,V_{i\,\cdots}=V_{\mu\,\cdots}+ 
\left(\p_\mu\phi^i+i\,[\,A_\mu,\phi^i\,]\,\right)\,V_{i\,\cdots}\,,& 
\label{napb}  \\[.2cm] 
& \cdots [\,\phi^i , \phi^j\,]\,V_{ij\,\cdots}\,,& 
\label{phicom}  \\[.2cm] 
&\ds V(X^\mu, X^i+\phi^i)= e^{\phi^i\p/\p X^i}V(X^\mu,X^i)\,.& 
\label{d-coup}
\eea 
The tensor $V_{\cdots}$ stands for $C^\prime_{M_1\cdots M_n}$ in the
Chern-Simons action and is given in terms of $E_{MN}=G_{MN}+B_{MN}$ or
the dilaton $\varphi$ in the DBI action. The first expression takes
the place of ordinary pull-back and is interpreted as its non-Abelian
generalization. Expressions of the second type appear in the BDI as
well as in the CS action where they result in a Dp-brane carrying
charges corresponding to larger branes. The third expression indicates
that the closed string backgrounds in the non-Abelian worldvolume
action should be regarded as functions of the non-Abelian scalars
\cite{Douglas:1995bn} (these cannot be seen via T-duality due to the
need for isometries).

It is easy to see that the above structures cannot follow from
covariant expressions on choosing a coordinate gauge: A covariant
action will contain all components of $\phi^M$, including $\phi^\mu$,
which are generically non-zero even in the static gauge, as discussed
in section \ref{ssCov}. Also, being matrix valued, the $\phi^\mu$
cannot be gauged away or reintroduced into the action by ordinary
coordinate transformations, unlike Abelian scalars. On the other hand,
these components of the non-Abelian scalars do not appear in
\rf{napb}-\rf{d-coup} which shows that they cannot follow from a
covariant expression on fixing a GCT gauge. Besides this, the
expressions \rf{napb}-\rf{d-coup} also contain other sources of
non-covariance involving the Christoffel connection. These contain
derivatives of the metric and arise at higher orders in perturbation
theory to which the derivation in \cite{Myers:1999ps} is not 
sensitive. 

The puzzle is that the above apparent inconsistency with general
covariance is not entirely a result of overlooking terms in the
calculation.  Rather, the T-duality used in \cite{Myers:1999ps} is
certainly valid for terms not containing derivatives of closed string
backgrounds (the $V_{M_1\cdots M_n}$ above). But the incompatibility
with general covariance already shows up at this level, within the
domain of validity of the derivation. This also prevents us from
adding the missing components $\phi^\mu$ by hand, without a deeper
understanding of their absence, as this would amount to changing by
hand the outcome of a valid derivation. However, terms involving
derivatives of the background fields (for example, the connection) are
missed by T-duality and can be added by hand if required for
covariance. One should then be able to reproduce them by a microscopic
calculation.
 
An understanding of the problem and a mechanism for its resolution
emerge from our worldsheet considerations. Recall that the worldsheet
boundary action obtained by T-duality in \rf{DpsTdna} involves the
same set of scalar field couplings as appear in \rf{napb}-\rf{d-coup},
and therefore exhibits the same inconsistency with covariance.  On the
worldsheet boundary the missing components $\phi^\mu$ could be
reinserted into the action by adding to it terms that vanish by virtue
of the boundary conditions. This restores covariance without affecting
the dynamics. The implication is that the complete worldvolume theory
should also have a corresponding symmetry (irrespective on how it is
implemented) that goes beyond GCT and allows us to insert into the
action the missing terms needed for covariance. The actual
implementation of this symmetry also follows from the worldsheet
theory. It involves a shift of the coordinates $X^\mu$ by matrix
valued fields the admissibility of which is not evident from the known
structure of the worldvolume action. The realization of this idea in
the Abelian theory was verified in the previous subsection. The
details and some fine tunning required in the non-Abelian case are
discussed below.  

An aspect not determined by the worldsheet considerations is when to
regard the scalar field as a normal vector $\phi^M$ and when to regard
it as a coordinate difference $\bar\phi^M$. Since the two differ by
higher derivatives of the metric, the T-duality used here and in
\cite{Myers:1999ps} (which has no $\alpha^\prime$ corrections) does
not distinguish between them. In fact, in \rf{napb}-\rf{d-coup},
$\phi^i$ seems to correspond to a vector whenever it appears within a
commutator and to a coordinate difference otherwise. On the
worldsheet, the difference does not show up since the shift is taken
to be infinitesimal, whereas $\bar\phi$ differs from $\phi$ at higher
orders.  The choice between the two has to be made depending on what is
consistent with covariance. In principle, all this can be verified by
microscopic calculations although that will not be attempted here.

Now we can demonstrate explicitly how the implementation of this
enlarged symmetry promotes \rf{napb}-\rf{d-coup} to covariant
expressions. First consider the expression $V(X^\mu, X^i+\phi^i)$ in 
\rf{d-coup}. We concentrate on the argument of $V$ since its tensor
index structure is part of \rf{napb}. Clearly the symmetry observed
on the worldsheet boundary allows us to shift $X^\mu$ infinitesimally
by $\phi^\mu$, resulting in $V(X^M+\phi^M)$. In analogy with the
Abelian case, covariance demands that the vector $\phi^M$ in the
argument should be replaced by the non-Abelian ``coordinate 
difference'' $\bar\phi^M$, given by the Riemann normal coordinate 
relation \rf{rnc}. On restricting to transverse scalars,
\beq
\bar\phi^M=\phi^M-\sum_{n=2}^\infty \frac{1}{n!}\Gamma^M_{L_1\cdots
L_n} \phi^{L_1}\cdots \phi^{L_n}\,.
\label{rncNA}
\eeq
Here we have simply generalized the Abelian expression to the
non-Abelian $\phi^M$ without bothering about the attendant
subtleties. The issue of normal coordinates involving matrices has
been considered in detail in \cite{DeBoer:2001uk} and will not be
discussed here further.  The function $V(X^M+\bar\phi^M)$ can be
expanded by the generalization of the Taylor expansion using the
normal coordinate formalism
\cite{Alvarez-Gaume:hn,Mukhi:1985vy,petrov},
\beq
V(X^M+\bar\phi^M)= e^{\phi^M\nabla_M} V(X^M)+ \cdots
\label{cd-coup}
\eeq
where the covariant derivative contains the Christoffel connection
$\Gamma^K_{MN}$ and the ellipses represent corrections involving the
curvature tensor. This is the covariant generalization of \rf{d-coup}. 

Next we consider the expression $V_{\mu\,\cdots}+D_\mu\phi^i\,
V_{i\,\cdots}$ \rf{napb}. The static gauge used can be slightly
generalized to $X^\mu=X^\mu(\xi^\al)$, $X^i=const$ for which the
derivation in \cite{Myers:1999ps} still goes through. We can also
regard the trace of $\phi^i$ (or part of it) as contributing to $X^i$.
This takes us out of the static gauge and leads to $\p_\al X^M
V_{M\,\cdots}+D_\al\phi^i\, V_{i\,\cdots}$. Simply shifting $X^\mu$ is
not enough to render this covariant. But recall that the correct
implementation of the shift transformation on the worldsheet resulted
in $X^\mu \rightarrow X^\mu+\phi^\mu$ and $\p_\al X^\mu \rightarrow
\p_\al X^\mu+D_\al\phi^\mu$. This was achieved by the formal trick
described around equation \rf{AphiU} and justified there: We first
make a gauge transformation $\phi^\prime=U^{-1} \phi\, U$ with $U$
given by the Wilson line \rf{U} evaluated along some fictitious path
on the worldvolume ending at $\xi^\al$. This sets $\mA^\prime_\al=0$
at $\xi^\al$, leading to $\p_\al X^M V_{M\,\cdots}+\p_\al\phi^{\prime
i}\,V_{i\,\cdots}$. Now, the invariance of the boundary couplings
under the {\it infinitesimal} shift $X^\mu\rightarrow X^\mu +
\phi^{\prime\mu}$ allows us to insert the missing $\phi^\mu$
components completing this to $\p_\al (X^M +\phi^{\prime M})
V_{M\,\cdots}$. For this to be sensible on the worldvolume,  
the vector $\phi^M$ should be replaced by the coordinate difference
$\bar\phi^M$ \rf{rncNA}. In practice, the difference should emerge
from higher order corrections in the microscopic computation of 
the worldvolume action, as argued in \cite{DeBoer:2001uk}. Now we have
the covariant generalization of \rf{napb},
\bea
U\,\p_\al (X^M +\bar\phi^{\prime M}) V_{M\,\cdots}\,U^{-1}
&=& U\left(\p_\al X^M + \nabla_\al\phi^{\prime M} 
+ \cdots \right) V_{M\,\cdots} \,U^{-1} \nonumber\\[.1cm]
&=& \left(\p_\al X^M + \nabla_\al^{(A)}\phi^M 
+ \cdots \right) V_{M\,\cdots}
\label{cnapb}
\eea
where $\nabla_\al^{(A)}\phi^M=\nabla_\al\phi^M+i\,[\mA_\al,\phi^M]$ is
gauge and GCT covariant, and we have used \rf{dphip} and \rf{dphib}.
The ellipses denote curvature dependent terms. The tensor $V_{\cdots}$
itself has a structure as in \rf{cd-coup}.

Last we consider \rf{phicom}. The shift in $X^\mu$ cannot be used
directly to covariantize this and even on the worldsheet the
corresponding commutator term in \rf{DpsTdna} was rendered covariant
by a shift in the fermionic worldsheet coordinates in \rf{nabshift}
and not $X^\mu$ itself. As yet it is not clear how the coordinate
shift should be generalized to also implement this aspect of the
worldsheet symmetry on the brane worldvolume. Nevertheless worldsheet
considerations have shown that such a generalization of the shift
should exist and convert \rf{phicom} into the corresponding covariant
expression,  
\beq
\cdots [\,\phi^M , \phi^N\,]\,V_{MN\,\cdots}\,.
\label{cphicom}
\eeq
The covariant version of the non-Abelian scalar couplings in
\cite{Myers:1999ps} can thus be obtained by replacing the structures 
\rf{napb}-\rf{d-coup} by \rf{cnapb}, \rf{cphicom} and \rf{cd-coup},
respectively.  

As such, the covariant expressions above could be easily guessed and
the worldvolume action modified accordingly without invoking the
worldsheet theory and it matrix-valued shift symmetry. However, since
the action in \cite{Myers:1999ps} is obtained following a consistent
procedure, there is no room for introducing into it new terms by hand
(except for those which fall beyond the scope of the derivation, like
terms involving the Christoffel connection). Therefore, to address the
issue or covariance of scalar couplings in \cite{Myers:1999ps}, an
approach like the one followed here becomes indispensable. It not only
yields a covariant expression, but also highlights the existence and
importance of matrix-valued coordinate transformations.

To summarize, the above discussion leads to the following picture of
the relation between GCT covariance, matrix-valued transformations and
T-duality in the worldvolume theory:  
\begin{itemize}
\item 
The complete non-Abelian worldvolume theory has an enlarged symmetry
group consisting of matrix-valued coordinate transformations (MCT)
which contains ordinary general coordinate transformations (GCT) as a
subgroup. The general structure of MCT and how it includes GCT has not
been specified. Issues pertaining to this type of enlarged symmetry
has been considered in \cite{DeBoer:2001uk,VanRaamsdonk:2003gj}.
\item
One can now fix a gauge using MCT. The Dp-brane action
obtained by T-duality in \cite{Myers:1999ps} is in fact in such a
gauge (explicitly, this involves fixing the static gauge for $X^M$ and
then using MCT to transform away the non-Abelian scalars $\phi^\mu$
keeping $\phi^i$). Since MCT contains GCT, gauge fixing the former
also fixes the latter. 
\item
A gauge that is fixed using MCT can only be undone by a matrix valued
transformation, and not by a GCT alone. Therefore, if we disregard
the possibility of matrix valued transformations, then expressions in
this gauge seem inconsistent with general covariance since they cannot
be obtained from covariant expressions by choosing a GCT gauge. 
\item 
This is the case with the non-Abelian worldvolume action of
\cite{Myers:1999ps}. The possibility of matrix valued coordinate
transformations is not evident from the action. However, worldsheet
considerations indicated the existence of a specific matrix-valued 
transformation which is the one needed to undo the MCT gauge fixing
and hence render the expressions GCT invariant.
\end{itemize}

One may regard the combination $X^M+\bar\phi^M$ or
$X^M+\bar\phi^{\prime M}$ as a matrix valued coordinate and
formulate MCT in terms of this. This is the approach followed in  
\cite{DeBoer:2001uk} in the context of D0-branes. Also see
\cite{VanRaamsdonk:2003gj} with emphasis on non-Abelian branes within branes.  
In this approach the understanding of ordinary general covariance
becomes more involved. Here we follow a more conservative approach of
treating $X^M$ and $\phi^M$ separately to make the GCT manifest. 

\subsection{Coupling to RR potentials through Clifford multiplication}

Before ending we will briefly comment on the coupling of charges
carried by D-branes to background RR-potentials, as contained in the
Chern-Simons part of the worldvolume action. It is well known that 
in the absence of the scalar field excitations a Dp-brane carries
charges, besides its own charge, corresponding to smaller branes on 
its worldvolume \cite{Douglas:1995bn}. These charges couple to
background RR-potentials through exterior multiplication of forms
which provides an elegant description of the couplings. 

The couplings get modified in the presence of non-Abelian scalars as
found in \cite{Taylor:1999gq,Taylor:1999pr,Myers:1999ps}. Physically
this implies that Dp-branes can also carry charges corresponding to
larger branes. The couplings of these new charges to RR-backgrounds
now also involve contractions and are no longer described by an
elegant exterior multiplication. It is desirable to find a unified
description of these couplings that replaces the exterior product. In
\cite{Hassan:2000zk} it was shown that such a unified description is
provided by the Clifford multiplication of forms. However, the details
of the formalism were not fully satisfactory: Clifford multiplication
can be regarded as a multiplication of Dirac gamma matrices. Since the
results in \cite{Myers:1999ps} were based on T-duality, it suggested
the use of gamma matrices associated with the T-duality group which do
not have a very natural meaning in the theory. Thus, while this
correctly reproduces the couplings, the formalism is tied to a specific
gauge.

It is much more appealing to base the Clifford multiplication on the
usual space-time gamma matrices that naturally occur in the theory.
These are also suggested by worldsheet considerations. However, their
use leads to the presence of extra terms in the action, not included in
\cite{Myers:1999ps}. One can check that these extra terms are
precisely the ones needed to complete the gauge fixed action
(involving couplings of the form \rf{napb}-\rf{d-coup}) to the one
consistent with covariance. As we have seen these terms in fact do
exist. Therefore a covariant description of the D-brane coupling to
RR-backgrounds is provided by Clifford multiplication as in
\cite{Hassan:2000zk}, but associated with the space-time gamma 
matrices.

\section{Conclusions}

In the first part of this paper (section \ref{sBVBV} to section
\ref{sWSBC}) we consider the N=1 open string worldsheet theory in general
non-constant backgrounds and give a unified description of D-brane
boundary conditions and boundary couplings in terms of a set of four
boundary ``vectors'', $\NX$, $\Npsi$, $\DX$ and $\Dpsi$. The complete
set of boundary couplings are constructed for the bosonic and N=1
Abelian and non-Abelian theories, consistent with supersymmetry,
T-duality and general covariance. One aspect of these couplings is
that, when written in a manifestly covariant form, the gauge field
always appears in the combination $A_M+\phi^LB_{LM}$. In the N=1 case,
the invariance of the boundary action under the NS 2-form gauge
transformation can be made manifest by adding to it a term 
$-\frac{i}{4}\Dpsi^\al\Npsi_\al$ that vanishes by virtue of the
boundary conditions. The presence of this term is crucial to insure
that the scalars $\phi^M$ can still be interpreted as infinitesimal
coordinate shifts. This also holds in the non-Abelian theory provided
the shift is correctly interpreted, for example in a gauge that sets
the gauge field to zero at the point under consideration.

One obvious use of these couplings is in the calculation of terms in
the worldvolume action, which is not the aim here. Rather we
investigate the behaviour of the scalar couplings under T-duality and
note that they behave exactly as in the Dp-brane worldvolume theory.
In particular, the non-Abelian scalars exhibit the same apparent
incompatibility with general covariance. However, on the worldsheet
boundary this has a simple resolution due to an invariance that allows
us to shift coordinates by appropriate matrices. It is then clear that
the same symmetry should also operate in the D-brane worldvolume
theory, rendering it consistent with general covariance, although its
existence is not evident from the known form of the action.

The resolution of the apparent inconsistency of non-Abelian scalar
couplings (as obtained by T-duality in
\cite{Taylor:1999pr,Myers:1999ps}) with general covariance is then
based on the following picture: The fully covariant non-Abelian
worldvolume action should also be invariant under a set of
matrix-valued coordinate transformations (MCT). The action obtained by
T-duality appears in a fixed MCT gauge, in which the components
$\phi^\mu$ of the scalars are gauges away. This looks incompatible
with covariance since this gauge fixing cannot be achieved or undone
by a general coordinate transformation. The resolution therefore lies
in the existence of MCT and the particular transformation required to
undo the gauge emerges from the worldsheet considerations.

In this paper we have taken the conservative approach of not combining
coordinates $X^M$ and scalars $\phi^M$ into non-Abelian coordinates,
as that would complicate the understanding of general covariance which
was our main concern. Once the minimum requirement of general
covariance is insured, one can take this approach and study the
problem on the lines of \cite{DeBoer:2001uk,VanRaamsdonk:2003gj}.

\vspace{.5cm}
\noindent{\Large{\bf Acknowledgment}}
\vspace{.3cm}\\
\noindent I would like to thank Dileep Jatkar and Suresh Govindrajan
for useful conversations. I would also like to thank Ulf Lindstrom,
Cecilia Albertsson and Maxime Zabzine for going through an early draft
of the paper and also bringing their result mentioned in footnote
\ref{f2} to my attention.  I am also grateful to the organizers of the
Second Crete Regional Meeting in String Theory for hospitality and the
opportunity to present this work.

\vspace{.3cm}
\newcommand{\resection}[1]{\setcounter{equation}{0}\section{#1}}
\newcommand{\appsection}[1]{\setcounter{equation}{0}\section*{Appendices}}
\renewcommand{\theequation}{\thesection.\arabic{equation}}
\appendix
\appsection

\resection{Supersymmetry conventions}
We use the following conventions for the N=1 supersymmetry on the
worldsheet parameterized by $\sigma^\pm=\frac{1}{2}(\tau\pm\sigma)$.  
The bosonic field $X^M$ and the fermionic fields $\phi^M_\pm$ are
combined into the superfield $X^M +\theta^+\phi_+ + \theta^-\psi_-
+ \theta^-\theta^+ F$. The supercovariant derivatives are 
$D_\pm =-\p/\p\theta^\pm + i\,\theta^\pm\p_\pm$. The supersymmetry
variations take the form,
\bea
&&\dels X^M = -\e^+\psi_+^M - \e^- \psi_-^M \,,\nonumber\\[.1cm]
&&\dels \psi^M_\pm = -i\,\e^\pm\p_\pm X^M \pm \e^\mp F^M \,,\nonumber\\[.1cm]
&&\dels F^M = -i\,\e^+\p_+\psi_-^M + i\, \e^-\p_-\psi_+^M\,.\nonumber
\eea
Now consider the superfield, ${\mathbb L}=E_{MN}(\bX)\bX^M\bX^N=
L_X+\theta^+L_++\theta^-L_-\theta^-\theta^+L_F$. From the
supersymmetry transformation of the F-term, we have  
$$
\dels L_F = -i\,\e^+\p_+L_- + i\, \e^-\p_-L_+ \,.
$$
The worldsheet action \rf{actionSF} is written as a sum of two parts
$S=S_\Si + S_{\p\Si}$. The first part involves the Lagrangian
density $\int d\theta^+ d\theta^-{\mathbb L} =L_F$ and therefore 
its supersymmetry variation is,
$$
\dels S_\Si =-i\int d\tau \left(\e^+ L_- + \e^- L_+\right)\br\,.
$$ 
Adding this to the supersymmetry variation of $S_{\p\Si}$ leads to 
$\dels S$ in \rf{susydS}.

\resection{Index Conventions} 

Capital letters from the middle of the alphabet $K,L,M,N$ label
10-dimensional space-time indices. $\al, \be$ denote worldvolume
indices and $\ha, \hb$ correspond to flat normal frame indices. In the
static gauge, the space-time coordinates identified with the brane
worldvolume coordinates are labeled by $\mu, \nu, \cdots$, while the
coordinate not along the worldvolume are labeled by $i, j, \cdots$.
The letters $a, b$ denote gauge group generators.

\par
\end{document}